\documentstyle[11pt,aasms4,epsf]{article}

\slugcomment{Accepted for publication in {\it The Astrophysical Journal}}

\begin{document}

\def\etal{{\it et al. }}

\title{HST Imagery and CFHT Fabry-Perot 2-D Spectroscopy in H$\alpha$ of the
Ejected Nebula M1-67: Turbulent Status\footnote{Based on
observations made with the NASA/ESA {\it Hubble Space Telescope}, obtained
at the Space Telescope Science Institute, which is operated by the
Association of Universities for Research in Astronomy, Inc., under NASA
contract NAS 5-26555. Also based on observations collected at the
Canada-France-Hawaii Telescope (CFHT), which is operated by CNRS of France, NRC
of Canada, and the University of Hawaii.}}

\author{Yves Grosdidier\altaffilmark{2,3,4,8} \&
Anthony F.J. Moffat\altaffilmark{2,3,5},\\ 
e-mail: yves@ll.iac.es, moffat@astro.umontreal.ca}

\and

\author{S\'ebastien Blais-Ouellette\altaffilmark{2,3,6},
Gilles Joncas\altaffilmark{7,3}, Agn\`es Acker\altaffilmark{4}}

\altaffiltext{2}{Universit\'e de Montr\'eal, D\'epartement de Physique,
C.P. 6128, Succursale Centre-Ville, Montr\'eal (Qu\'ebec) H3C 3J7,
Canada.}
\altaffiltext{3}{Observatoire du Mont M\'egantic, Canada.}
\altaffiltext{4}{Observatoire Astronomique de Strasbourg, UMR 7550,
11 rue de l'Universit\'e, F-67000 Strasbourg, France.}
\altaffiltext{5}{Killam Fellow of the Canada Council for the Arts.}
\altaffiltext{6}{Institute of Geophysics \& Planetary Physics, L-413,
Lawrence Livermore National Laboratory, 7000 East Avenue, Livermore CA,
94550, USA.}
\altaffiltext{7}{Universit\'e Laval, D\'epartement de Physique, Pavillon
Alexandre-Vachon, Sainte-Foy (Qu\'ebec) G1K 7P4, Canada.}

\altaffiltext{8}{Present address: Instituto de Astrof\'{\i}sica de Canarias, Calle V\'{\i}a L\'actea s/n, E-38200 La Laguna (Tenerife), Spain.}

\begin{abstract}

Bright circumstellar nebulae around massive stars are potentially useful to
derive time-dependent mass-loss rates and hence constrain the evolution of the
central stars. A key case in this context is the
relatively young ejection-type nebula M1-67 around the runaway
Population I Wolf-Rayet star WR124 (= 209 BAC), which exhibits a WN8 spectrum. With HST-WFPC2 we have obtained a deep, H$\alpha$ image of M1-67.
This image shows a wealth of complex detail which was briefly presented previously by
Grosdidier \etal (1998). With the interferometer of the Universit\'e Laval
(Qu\'ebec, Canada), we have obtained complementary Fabry-Perot H$\alpha$ data
using CFHT MOS/SIS.\\
From these data M1-67 appears more-or-less as a spherical (or elliptical,
with the major axis along the line of sight), {\it thick, shell} seen
almost exactly along its direction of rapid spatial motion away from the
observer in the ISM. However, a simple thick shell by itself would
not explain the observed multiple radial velocities along the line of sight. This velocity dispersion leads one to
consider M1-67 as a thick {\it accelerating} shell.
Given the extreme perturbations of the velocity field in
M1-67, it is virtually impossible to measure
any systematic impact of
the present WR (or previous LBV) wind on the nebular structure.
The irregular nature of the velocity field is 
likely due to either large variations in the density distribution of the ambient ISM,
or large variations in the central star mass-loss history.
In addition, either from the
density field or the velocity field, we find no clear evidence for a bipolar
outflow, as was claimed in other studies.\\
On the deep H$\alpha$ image
we have performed continuous wavelet transforms to isolate stochastic
structures of different characteristic size and look for scaling laws.
Small-scale wavelet coefficients show that the density field of M1-67 is remarkably structured in chaotically (or possibly radially)
oriented {\it filaments} everywhere in the nebula.
We draw
attention to a short, marginally inertial range at the smallest scales
(6.7--15.0 $\times 10^{-3}$ pc), which can be
attributed to turbulence in the nebula, and a strong scale break at larger
scales. Examination of the structure functions for different orders shows that
the turbulent regime may be intermittent.\\
Using our Fabry-Perot interferograms, we also present an investigation of
the statistical properties of fluctuating gas motions using structure
functions traced by H$\alpha$ emission-line
centroid velocities. We find that there is a clear correlation at
scales 0.02--0.22 pc between the mean quadratic differences of radial velocities and distance
over the surface of the nebula. This implies that the velocity field shows an
inertial range likely related to turbulence, though not coincident with the small inertial
range detected from the density field. The first and second order moments of the velocity increments are found to scale as $\langle|\Delta v(r)|\rangle\sim r^{0.5}$ and $\langle|\Delta v(r)|^2\rangle\sim r^{0.9}$. The former scaling law strongly suggests that supersonic, compressible turbulence is at play in the nebula, on the other hand, the latter scaling law agrees very well with Larson-type laws for velocity turbulence. Examination of the structure functions for different orders shows that
the turbulent regime is slightly intermittent and highly multifractal with universal multifractal indexes $\alpha \approx 1.90$--1.92 and $C_1 \approx 0.04 \pm 0.01$.
\end{abstract}


\section{Introduction}

\subsection{Nebulae Ejected by Hot Stars}
Nebulae around central stars of planetary nebulae (CSPN)
or Population I Wolf-Rayet (WR) stars have proven to be a powerful
diagnostic tool to understand how mass is lost in the post main-sequence
phases of stellar evolution. CSPN and some ($\approx 1/3$;
Marston 1995, 1996, 1999) massive WR stars are surrounded by
circumstellar nebulae (CN) made up of stellar material ejected during
successive stellar wind phases. On the whole, these nebulae often
appear to have similar physical properties, which point to a
common formation mechanism (Chu 1993) independent of the strong differences
between both types of hot stars.\\
The dynamics of planetary nebulae and CN surrounding Population I WR stars are
extremely
sensitive to the stellar wind velocity and mass-loss history.
In the context of the interacting stellar wind model, a spherically
symmetric, fast, hot stellar wind ($v_{exp}\sim10^3$ km s$^{-1}$) is catching
up and colliding with a pre-existing slower
($v_{exp}\sim10^{1-2}$ km s$^{-1}$), denser wind (CSPN: Kwok {\it et al.} 1978;
Chu \etal 1991; Balick 1994; Pop. I WR stars: Marston 1995, 1996, 1999;
Garc\'{\i}a-Segura \& Mac Low 1995; Garc\'{\i}a-Segura, Mac Low \& Langer 1996, and
references therein). The non-circular shapes of many CN suggest that the fast
wind is sweeping up a slow wind that is flattened by rotation or binary
effects (Garc\'{\i}a-Segura \& Mac Low 1995; Mellema \& Frank 1995). However,
a flattened, fast wind blowing out a spherically symmetric
slow wind can also lead to a non-circular CN (Frank \etal 1998).\\
Rayleigh-Taylor instabilities are expected to occur
as a consequence of such interaction, as well as other hydrodynamical
instabilities which may lead to turbulence in the nebula.  On a
{\em qualitative} basis, the interacting stellar wind model agrees with the
observations, but it is still difficult to determine the precise initial,
overall geometry of the colliding winds. In particular, the question arises
as to the impact of pre-collision, apparently {\it universal}, clumping
or possible turbulence of the {\it hot}-star winds (Robert 1992; Balick
{\it et al.} 1996;
L\'epine \etal 1996; Acker \etal 1997;
Grosdidier {\it et al.} 1997; Eversberg \etal 1998;
L\'epine \& Moffat 1999; Grosdidier \etal 2000, 2001;
and references therein)
on the dynamics/morphology of CN. Moreover,
despite preliminary work done by Stone, Xu \& Mundy (1995), the incidence
of a highly variable (as well in the spatial as in the time domain)
mechanical luminosity
[$\propto \dot{M}(r,\theta,\phi,t)\times v_{flow}^2(r,\theta,\phi,t)$] still
has to be investigated in detail. In particular, the turbulence of the
ejected nebulae has to be studied in detail in order to test whether
it is related to pre-existing turbulence (or any other process of
fragmentation, hence variability) that already arose in the {\it winds}
themselves before they interacted.\\
Recall that the spectral variability
originating in massive WR stars and CSPN shows the WR phenomenon to be
remarkably similar in both (Grosdidier \etal 2000, 2001).
Despite the strong differences between both types of hot stars,
this supports the understanding of the WR spectral phenomenon as being purely
atmospheric. This, in line with the believed similarity of the
processes that are at play in the formation of the related CN (Chu 1993),
places one in a position to check also for universality of the shaping of
CN surrounding {\it any} hot stars.

\subsection{Turbulence in Ejected Nebulae and HII Regions}
Like HII regions, gas motions in nebulae ejected by hot stars are
characterized by very large Reynolds numbers. For example, the Reynolds
numbers exhibited by typical HII regions may be as large as $10^9$ (Boily 1993), using 1 pc for
the characteristic spatial scale, 10 km s$^{-1}$ for the velocity scale and considering the
normal kinematic viscosity for ionized hydrogen gas at electronic temperatures of about 10,000 K. Such values are well
above the critical values ($10^2$--$10^4$) determining the transition to 
turbulence. Like HII regions, it is expected that
the kinematics of nebulae ejected by hot stars would be best
described in terms of turbulent flows. Turbulence is characterized by
a high degree of non-linearity in the governing equations and complex
coupling mechanisms, leading to a variety of physical/transport processes
occuring over a great variety of scale lengths.\\
Turbulence is inherently {\it apparently} stochastic, since the flow
variables ({\it e.g.} velocity, pressure, density) fluctuate in an unpredictable
manner about their mean values. However, turbulence also exhibits specific
length scales and scaling laws, with rapidly increasing numbers of smaller
features as a result of energy cascading down a ladder leading ultimately
to dissipation (Fleck 1996). At small scales, turbulent energy returns to the
ambient medium as thermal energy. The nearly dissipationless energy cascade makes the flow
not strictly unpredictable
if one attempts to study quantitatively its structure via a statistical
approach. Indeed, the signature of turbulence is confirmed by the presence of
correlations in the flow variables between neighboring points that can be detected by means of
statistical functions (Scalo 1984). The non-linearity of the equations
of hydrodynamics causes high amplitude structures to change their
shape in a way that continuously broadens the wave spectrum to include
smaller and smaller scales. A simple, unique wave train is expected to
steepen into a shock front damping at the leading edge. However, an intricate
pattern of motions likely drives a cascade of energy dissipation
taking on a fractal structure (Henriksen 1994, and references therein).\\
A statistical search for any degree
of correlation between neighboring points could be carried out via two main
approaches: autocorrelation functions or structure functions (these two statistical
tools are more reliable than the dispersion-scale
relation; Scalo 1984, Miville-Desch\^enes \& Joncas 1995). The so-called `structure function'
is defined as the average of the quadratic
differences of velocities as a function of the spatial separation
between the points where the velocity is measured (Miesch \& Bally 1994). In this
study
we favor the structure function calculation because it is more robust than the 
autocorrelation method. The robustness of the structure function technique originates
in one important fact: its application requires the use of velocity
{\it differences}
rather than velocity `coherence' (in the case of the autocorrelation method) 
between points separated by a given distance. Indeed, the use of the autocorrelation
function assumes that the statistical properties of the flows are not a function
of position, the velocity and density fields being considered {\it stationary}.
When using structure functions, we assume a less stringent hypothesis:
{\it local stationarity}. However, note that at large scales the flows always
lose their stationarity (Scalo 1984).
The aim of the present paper is to describe the turbulent behavior of nebulae
ejected by hot stars, both from the density structure (via high-resolution imagery)
and the kinematical structure (via high spatial resolution Fabry-Perot
interferometry). The search for turbulence will be carried out through the
calculation of the moments of the density/velocity increments for different orders
({\it i.e.} we will not limit our investigation to the second order structure function, as is generally the case in other studies). Examination of the
structure function for different orders additionally provides a way to estimate
the level of intermittency in the flows.\\
M1-67, surrounding the WR star WR124 (WN8), was chosen for its
relatively large apparent angular diameter (which allows one to study any possible
energy
cascade within a large span of scale lengths) and its {\it suspected} weak interaction
with the ambient interstellar medium. The latter condition ensures that 1) the
ejecta would be less affected by collision and mixing with the pre-existing ambient
gas, and therefore 2) the dynamics and structure of the ejected nebula would
best reflect the initial conditions of the ejected winds. This, in line with the
apparently universal, ubiquitous clumping of hot stellar winds, places one in
a position to test the possible impact of the wind fragmentation on the nebular
fragmentation. Note that high resolution optical spectra of WR124 recently obtained by Acker (priv. comm.) at the Observatoire de Haute Provence (France) and Marchenko \etal (1998) results indeed show the central star of M1-67 to be variable.

\section{The Nebula M1-67 and its Central Star WR124}
\begin{figure}
\epsfxsize=17cm
\epsfysize=17cm
\epsfbox{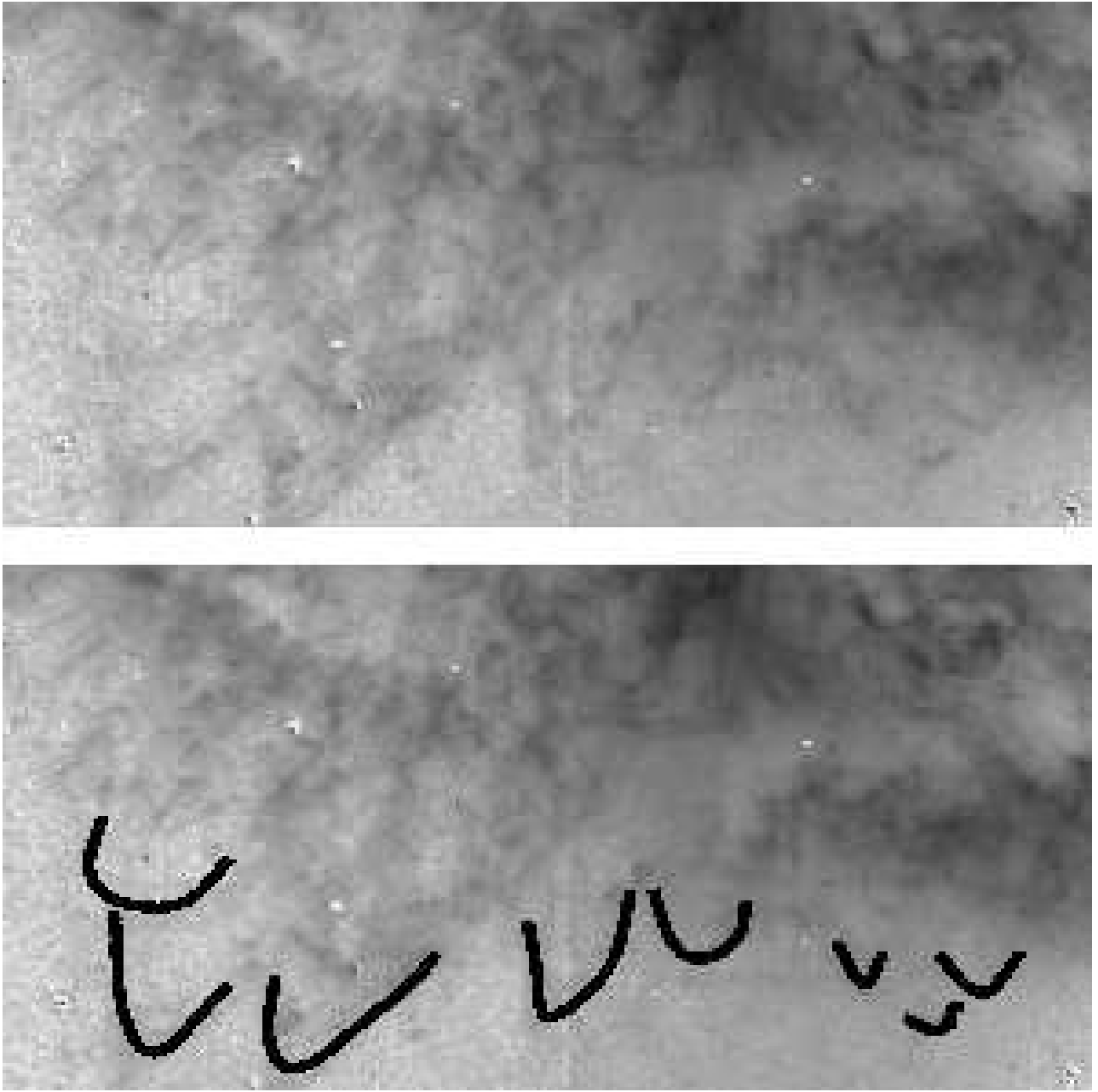}
\caption{Enlargement of the faint outer north-eastern region ($PA \approx 10^{\circ}$--30$^{\circ}$) of the 
nebula (upper panel) where at least 9 `reversed' bow-shocks are visible (see
lower panel). Such features are seen throughout the nebula although better
detected at the periphery. A few residuals of subtracted field stars also
appear. The chosen contrast emphasizes the dimmer features.
\label{bowshocksm167}}
\end{figure}

Ground-based, narrowband images and discussion of the ejected
nebulosity M1-67 surrounding the cool nitrogen WN8 WR star WR124 (= 209
BAC) are given in Esteban {\it et al.} (1991), Esteban {\it et al.} (1993),
Sirianni {\it et al.} (1998), Crowther \etal (1999),
and references therein. Until 1991 the central star of M1-67 was often
mistaken for a CSPN. From a study of the Na I D$_2$ interstellar
absorption line in the context of Galactic rotation, Crawford \& Barlow (1991)
showed that the distance
to WR124 is 4--5 kpc. For such a large distance, the central star
must be a massive WR star rather than a CSPN, given its relatively high
apparent brightness, $V\approx 11.6$ (Massey 1984), and the huge interstellar
reddening to M1-67 ($E_{B-V}\approx 0.9$--1.5; Solf \& Carsenty 1982; Esteban \etal
1991; Crowther \etal 1995b). This is compatible with $M_V \approx -6.0$,
which is typical for WN8 stars. Photo-ionization modelling of M1-67 has
been made by Crowther \etal (1999), who were able to trace the central star
H-Lyman continuum flux distribution, which is normally not directly
detectable. The main outcome of this work was to establish the crucial
importance of a line-blanketed energy distribution for proper modelling of
the WN8 atmosphere.\\
As an E-type (ejected material) nebula (see Chu, Treffers \& Kwitter 1983;
Smith 1995 and references therein), M1-67 constitutes an object in
its earliest phase of wind interaction, when the ejecta are least affected
by collision and mixing with the interstellar medium, and thus best reflect the
initial conditions of the ejected winds. This is supported by detailed
abundance analysis of the nebula: N/O $\approx$ 2.95 (Esteban {\it et al.} 1991),
rather than N/O $\approx$ 0.07 for the Galactic ISM (Shaver {\it et al.}
1983). Note that the E-type Galactic CN RCW58 (central star: WR40; spectral type:
WN8), is also obviously enriched and
must contain stellar ejecta; however this nebula already shows a circumstellar
ring which suggests a more pronounced interaction with its ambient medium.
For the same reason, we do not consider the multiple ring (Marston 1995) CN
surrounding the WN8 star WR16: the two outer rings are likely E-type
ring nebulae, but their dynamical ages are relatively long (some $10^6$ yrs
and $7\times 10^5$ yrs). In addition, the inner ring is likely
a W-type (wind-blown bubble; see Chu {\it et al.} 1983 and references
therein) CN.\\
Ground-based observations ({\it e.g.} Chu \& Treffers 1981; Sirianni {\it et al.} 
1998) exhibit a nearly spherical, possibly bipolar, patchy structure made up of many
bright unresolved knots and filaments. The basic properties of M1-67 are: diameter 110--120''
 ({\it i.e.} 2.4--2.6 pc at a distance of 4.5 kpc;
Grosdidier \etal 1998), mean expansion velocity 40--45 km s$^{-1}$ (Sirianni
{\it et al.} 1998), implying a dynamical age of the order of a few $10^4$ yrs.\\
In the framework of our search for the direct influence of a clumpy, turbulent-like
stellar wind on the surrounding nebula it ejects and more generally on the
interstellar medium (Moffat 1998), we have obtained {\it Hubble Space Telescope}
(HST) {\it Wide Field Planetary Camera Two} (WFPC2) images of the nebula
M1-67.
Figure \ref{imagem167} is an enlargement of the picture reproduced in
Grosdidier {\it et al.} (1998). They reported no overall global shell
structure to the nebula (see also Grosdidier \etal 1999) from the projected
radial distribution of the azimuthally-averaged H$\alpha$ surface brightness.
Rather, the radial profile centered
on WR124 agrees with a decreasing, wind-density power-law distribution with increasing
distance from the central star. Grosdidier {\it et al.} (1998) understood the absence
of a well defined shell as evidence for an extremely small age of the
nebula: the WR phase may have just turned on, making the interaction zone
({\it i.e.} the
nebula) of the present fast wind with the previous, slower wind still undetectable
(which is in line with the presence of H emission lines in the spectrum of WR124;
Crowther \etal 1995a).\\
However, these deep HST/WFPC2 images in H$\alpha$ of the nebula also suggested that
M1-67 may likely
be the imprint of a previous, luminous blue variable (LBV) slow wind ejected from
the WR central star
WR124, which is supported by other arguments
(Sirianni {\it et al.} 1998, and references therein; Massey \& Johnson 1998) as
opposed to a classical red supergiant very slow wind (Smith 1996) blown and compressed by
the present WR wind. Note that ubiquitous {\it reversed} bow-shock like structures
are detected throughout the nebula (the more prominent ones being detected at
the outer edge of M1-67, $PA\approx$ 10$^{\circ}$--20$^{\circ}$ and 130$^{\circ}$--160$^{\circ}$; see Figure \ref{bowshocksm167}).
Generally, the whole periphery of the nebula appears made up of numerous small (1--5'')
reversed bow-shock like structures. These features
can be interpreted as the result of the thin-shell instability of the highly variable
previous LBV stellar wind interacting with {\it itself} ($\it e.g.$ Stone {\it et al.} 1995). In addition, some dense, persisting clumps have
possibly been ejected directly from the LBV stellar surface (Grosdidier {\it et al.}
1998). 

\section{Observations and Data Reduction}

\subsection{HST Imagery}
Here, we expand on the description given in Grosdidier {\it et al.} (1998).
Four WFPC2 images of M1-67 with a total combined exposure of 10,000
seconds were taken in March 1997 with the HST
through the narrowband F656N H$\alpha$ filter (Biretta \etal 1996). They were
combined (with the task GCOMBINE in STSDAS/IRAF\footnote{IRAF is distributed
by the National Optical Astronomy Observatories, operated by the
Association of Universities for Research in Astronomy, Inc., under
cooperative  agreement with the National Science Foundation.}) to
produce a single deep H$\alpha$ image cleaned from bad/hot pixels
and cosmic rays. Before the combination of the four F656N H$\alpha$ images, we
applied proper shifts to them in order to achieve alignments to within
$\pm$0.01--0.03 pixels. The shifts were determined using at least
15 (Planetary Camera; PC) and 25--30 (Wide Field Cameras 1, 2 and 3; WFs)
bright stars and calculating the relative displacements between each exposure
through the estimation of the star centroids.
Note that rather than applying one shift to an image to match the other,
we applied half-displacements to both images. This
reduces the differences in artificial broadening of the stellar profiles
between the two images.
Finally, we achieved full widths at half maximum spanning 2.4--2.5 pixels ($\approx 0.11$'') for
the PC, and 1.7--2.0 pixels (0.17--0.2'') for the WFs, in the H$\alpha$ band.\\
The four narrowband F656N H$\alpha$ images were straddled in wavelength by
broadband V and R images: four WFPC2 images of the same
field taken through each of the broadband F675W R and F555W V filters (Biretta \etal
1996) were separately combined in the same way to produce similar high
signal-to-noise ratio images of `continuum' starlight relatively close
to the wavelength of H$\alpha$. The V and R images enabled us to {\it interpolate}
the stellar light to the wavelength corresponding to H$\alpha$. 
After proper flux scaling and spatial shifting (also within $\pm$0.01--0.03
pixels), the interpolated broad-band image\footnote{The V-band component
being almost negligible.} was carefully subtracted from the
H$\alpha$ image in order to produce a deep continuum-subtracted H$\alpha$
image with the field stars removed.
Note that we achieved full widths at half maximum spanning 2.4--2.5 pixels for
the PC, and 1.7--2.0 pixels for the WFs, both in the R and V bands. These
values are similar to those obtained for the combined H$\alpha$ image and
permited an optimum continuum starlight subtraction.\\
The scaling in intensity between the interpolated broad-band image and the 
combined H$\alpha$ image was made by measuring the flux of at least
20 bright stars in each chip, and minimizing
the  residuals in the continuum-subtracted H$\alpha$ image. This procedure
works quite well for H$\alpha$ line-free stars, which are the
majority. For WR124 we applied a different flux scaling because this WN8 star
emits a copious amount of H$\alpha$+HeII$\lambda$6560 light. It is worthy of
note that we have obtained all images within a contiguous span of time (6
HST orbits). This reduced problems with potential variable stars in the
subtraction of starlight from the original H$\alpha$ image.
Note that the images were originally pipeline processed before being released
but required a full recalibration because of severe changes in the dark
files.

\subsection{Fabry-Perot Interferometry}
Nebular kinematics can be studied very effectively
by observations at optical wavelengths with the use of a scanning Fabry-Perot
interferometer in combination with a two-dimensional detector.
This technique allows one to probe
the velocity field with a wider field of view and higher throughput than with
a normal long-slit, high-dispersion spectrograph.\\
With the servo-stabilized Fabry-Perot (FP) interferometer of the Universit\'e
Laval (Joncas \& Roy 1984), we obtained complementary FP H$\alpha$ data
of M1-67 using
MOS/SIS\footnote{http://www.cfht.hawaii.edu/Instruments/Spectroscopy/SIS/}
(Le F\`evre \etal 1994) at the Canada-France-Hawaii Telescope (CFHT), in
1996 August 30.
The instrument was mounted at the
Cassegrain $f/8$ focus and covered a field of about 200'' on a side.
An H$\alpha$ narrow-band filter ($\delta\lambda=8.8$ \AA) centered at
6575 \AA\  was used as a pre-monochromator. The
combination of tilting and air temperature blue-shifted the central wavelength
of the filter to match the red-shifted H$\alpha$ line originating in M1-67
(mean peculiar velocity: about +137 km s$^{-1}$; Sirianni {\it et al.} 
1998).
The FP interferometer has a {\em finesse} $F=30$ and a free spectral range of
392 km s$^{-1}$ at H$\alpha$, where the interference order is
$p=765$. Since both the interferometer and filter were tilted, we
avoided ghost images (Georgelin 1970). We used the 2048 $\times$ 2048
pixel STIS2 CCD (gain = 4 e$^-$ ADU$^{-1}$, readout noise = 8 e$^-$, pixel
width = 21 $\mu$) with a spatial sampling of 0.30'' per pixel. We
employed the MOS/SIS tip/tilt image stabilizer in order to improve image
quality and achieved effective seeing spanning the range FWHM
$\approx$ 0.6--0.7''. The useful data cube was 1500 $\times$ 1500
pixels in the spatial domain, and 66 channels ($2.2\times F$, in order to
satisfy the Nyquist criterium) in the wavelength domain, giving a velocity
sampling of 5.9 km s$^{-1}$ per channel. The integration time
was 200 sec channel$^{-1}$. A neon calibration lamp was used to obtain rest
frame interferograms. Finally, the data processing was done using the IRAF and
ADHOC\footnote{http://www-obs.cnrs-mrs.fr/ADHOC/adhoc.html} (see
Amram \etal 1995, Blais-Ouellette \etal 1999) packages.

\section{The Turbulent Status of M1-67}

\subsection{HST Imagery: Wavelets for Structure Function Analysis}
\subsubsection{Method\label{method}}
We favor the use of wavelets to perform structure function calculations
because this method is much more robust than the classical study of the scaling
behavior via the power spectrum estimation as a function of the modulus of
the wavevector $k$. Indeed, the latter method is sensitive to biases in the
data, {\it e.g.} smooth polynomial trends, that can easily be shown to drastically affect the power spectrum scaling law. In addition, this method is appropriate when the
fractal signal under consideration is not associated to a unique roughness
exponent $H$ (Schmittbuhl \etal 1995).\\
Using 2D wavelets, we have analyzed our deep continuum-subtracted
H$\alpha$ image of M1-67 to isolate stochastic structures of different
characteristic size. Hence we effectively looked for scaling laws, as has
been clearly demonstrated in other astrophysical contexts ({\it e.g.}
clumping in CO: Gill \& Henriksen 1990; clumping in WR winds: L\'epine 1994).
In what follows, the deep H$\alpha$ image will be considered as a two-dimensional
function noted $f$.\\
Following Muzy \etal (1993), we have explicitly used wavelets to perform a
structure function analysis. We have considered the field of `increments' of
$f$ over a projected spatial distance $r$ at the pixel $(x,y)$, 
\begin{equation}
\Delta f(r;(x,y))=f((x,y)+R/2)-f((x,y)-R/2),
\end{equation}
by the continuous wavelet transform of
$f$ at the scale $r$, $T_{\Psi}[f](r;(x,y))$,
\begin{equation}\label{eqdef}
\Delta f(r;(x,y))\approx T_{\Psi}[f](r;(x,y))=r^{-2}\times[f\ast \Psi_r](r;(x,y)),
\end{equation}
where $R$ is a two-dimensional increment vector (on the plane of the sky) of
length $r$, $\ast$ denotes
a convolution and $\Psi_r$ is the analyzing wavelet at the spatial scale $r$
(Muzy \etal 1993).
In our study, the so-called `mother-wavelet', $\Psi$, is the  2D `Mexican hat'
(formally the second spatial derivative of $e^{-(x^2+y^2)/2}$:
$(2-x^2-y^2)e^{-(x^2+y^2)/2}$) and
$\Psi_r(x,y)=\Psi(\frac{x}{r},\frac{y}{r})$ (Farge 1992). It is worth noting
that the Mexican hat wavelet is orthogonal to polynomials of orders 0 and 1. Therefore, the wavelet transform is insensitive to linear trends, hence linear non-stationary components, in the signal.\\
The statistical moments
of $|\Delta f(r;(x,y))|$ (the two-point correlation known as `structure
function of order $p$') were estimated through the statistical
moments of $|T_{\Psi}[f](r;(x,y))|$ by averaging over the location:
\begin{equation}\label{eqconj}
\langle|\Delta f(r)|^p\rangle \approx C \int|T_{\Psi}[f](r;(x,y))|^p\,dx dy,
\end{equation}
where $C$ is an arbitrary constant.
Note that Muzy \etal (1993) applied this formalism to 1D signals. Here, we
generalize their approach to 2D signals.
In section \ref{tests} we will perform tests on artificial fractal
2D signals, and show that this procedure is correct.
Since this approach does not depend on the analyzing wavelet (Muzy \etal
1993), one is in a position to test for scaling laws of the form:
$\langle|\Delta f(r)|^p\rangle\sim r^{\zeta(p)}$.\\
Note that $\zeta(0)=0$ by definition and $\zeta(p)$ is a smooth,
differentiable, monotonically non-decreasing, concave function of $p$ (if
the signal has absolute bounds), no matter how rough the data are
({\it e.g.} Marshak \etal 1994). Continuous signals satisfy
$\langle | \Delta f(r) | \rangle\sim r$, hence $\zeta(p)=p$. Processes
with $\zeta(p)\propto p$ are called `mono-affine' or `monofractal', whereas
processes with variable $\zeta(p)/p$ are called `multi-affine' or
`multifractal'.

\subsubsection{Tests on Artificial 2D Fractal Signals\label{tests}}
\begin{figure}
\epsfxsize=14cm
\epsfysize=14cm
\epsfbox{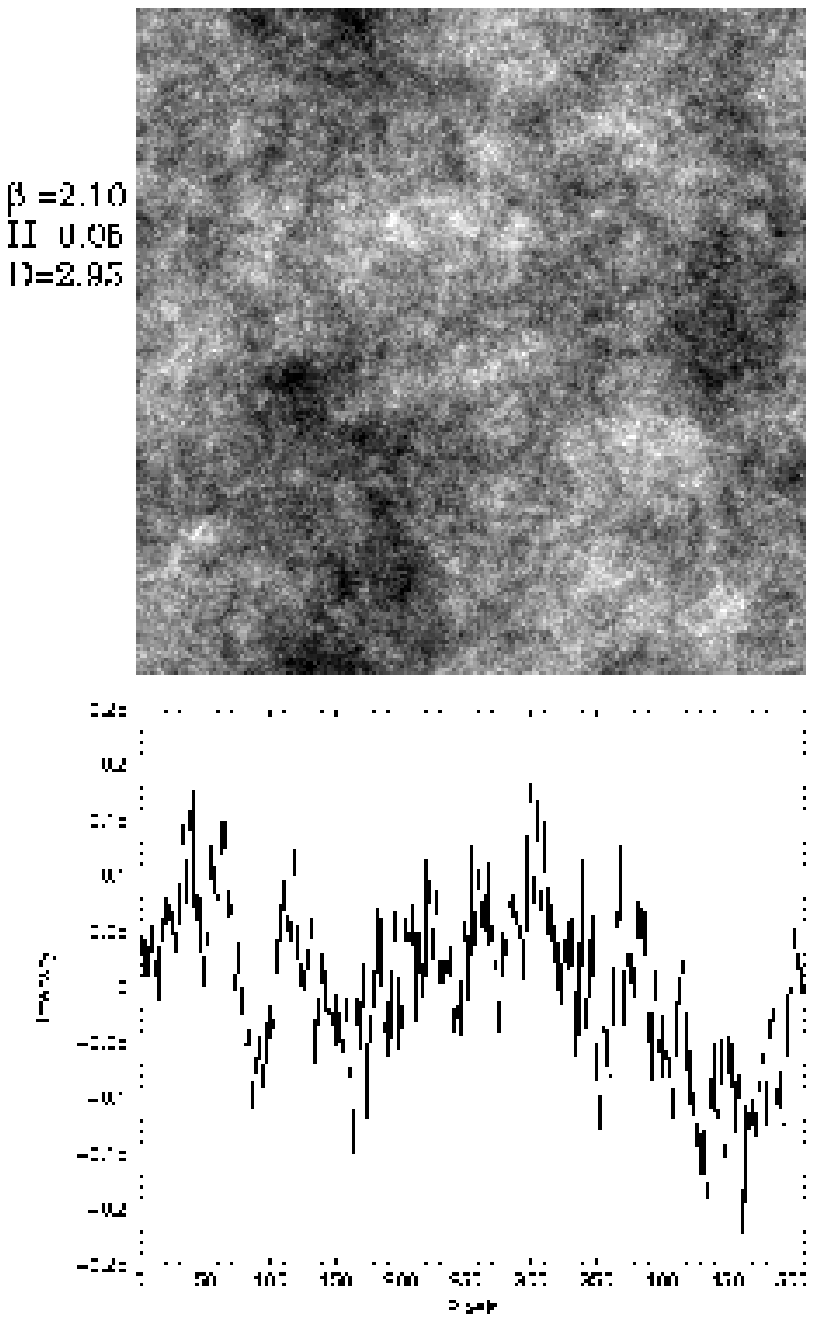}
\caption{Artificial fractal signal with $H=0.05$ (upper panel), along with a horizontal
cut in the middle of the image (lower panel).\label{artif1}}
\end{figure}

\begin{figure}
\epsfxsize=14cm
\epsfysize=14cm
\epsfbox{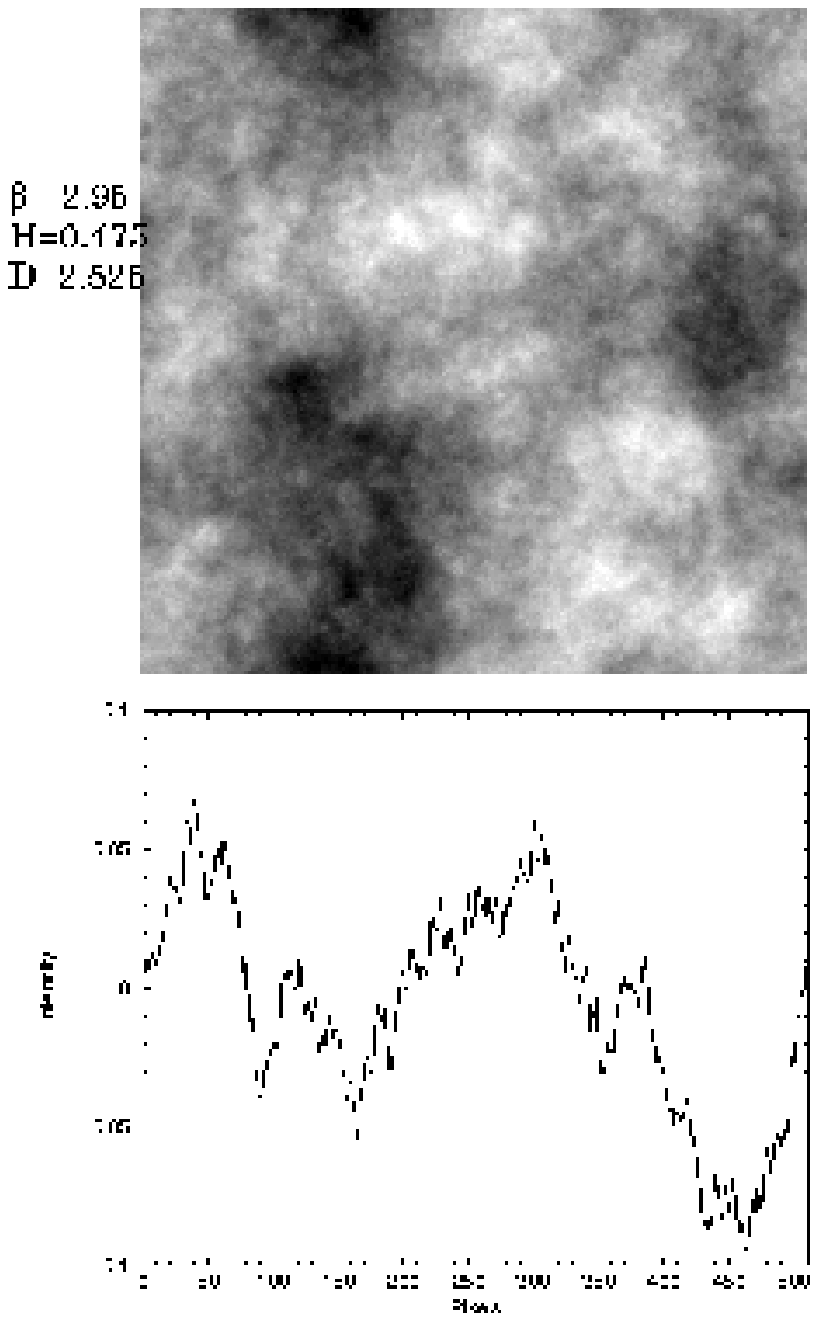}
\caption{Artificial fractal signal with $H=0.475$ (upper panel), along with a horizontal
cut in the middle of the image (lower panel).\label{artif2}}
\end{figure}

\begin{figure}
\epsfxsize=14cm
\epsfysize=14cm
\epsfbox{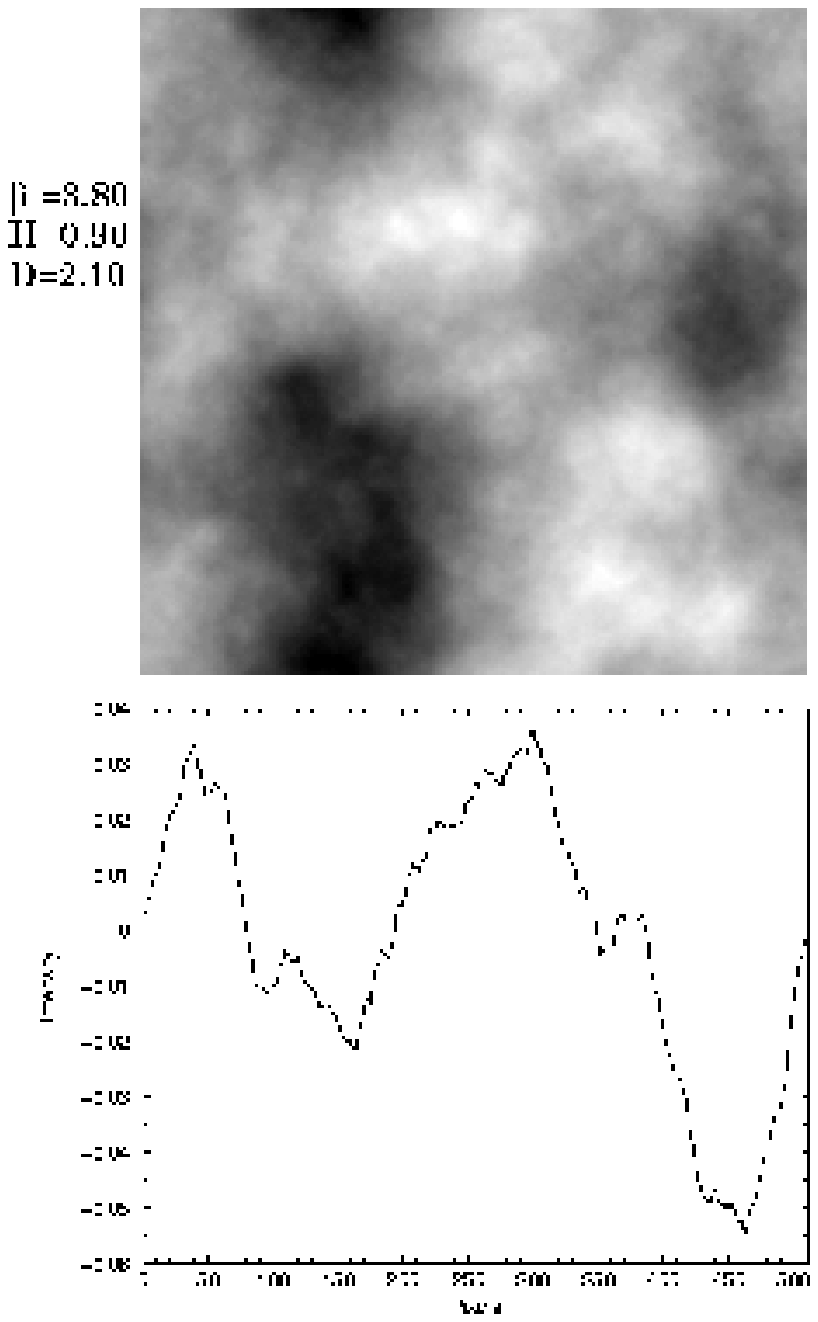}
\caption{Artificial fractal signal with $H=0.90$ (upper panel), along with a horizontal
cut in the middle of the image (lower panel).\label{artif3}}
\end{figure}

\begin{figure}
\epsfxsize=14cm
\epsfysize=14cm
\epsfbox{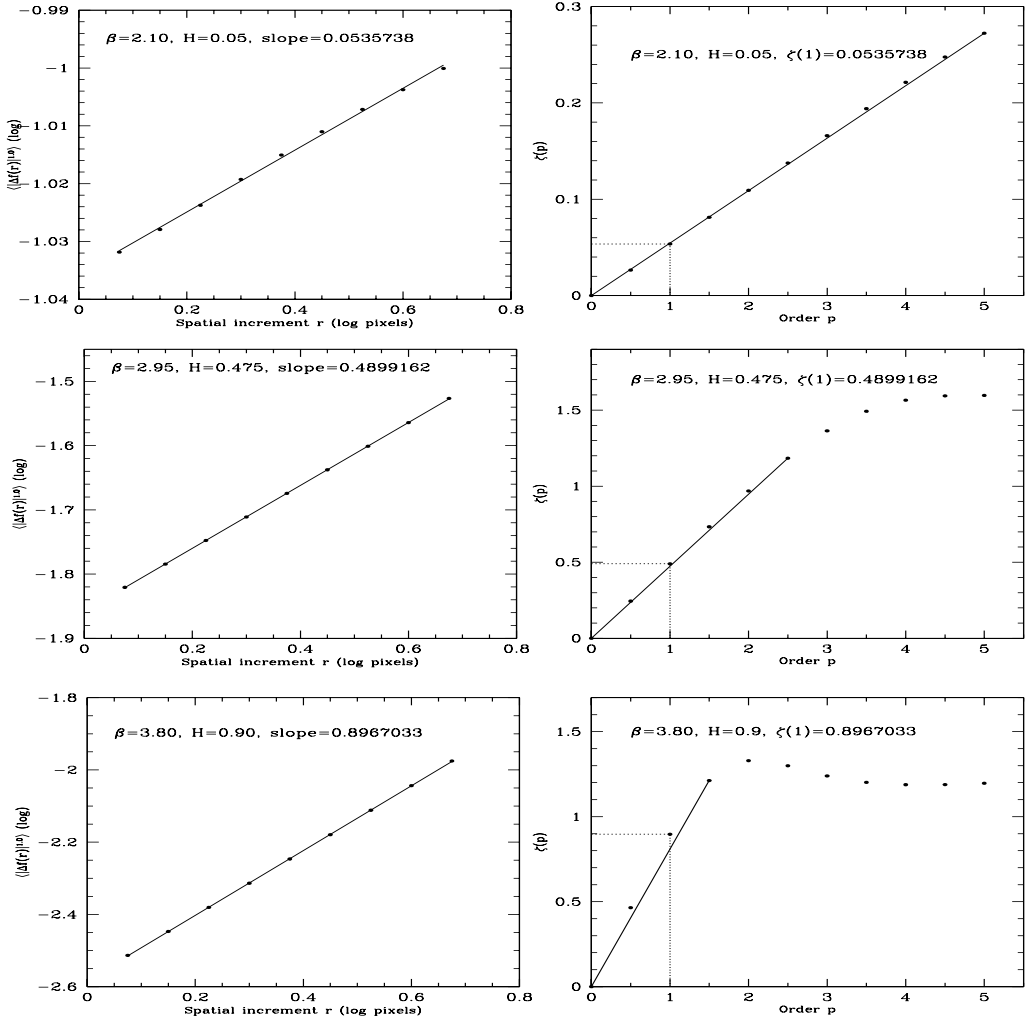}
\caption{{\it Left panels:} Persistence parameter retrieval for $H=0.05$, $H=0.475$ and $H=0.9$ (cf. Figures \ref{artif1}, \ref{artif2} and \ref{artif3}). {\it Right panels:} Structure function analysis of the signals shown in Figures \ref{artif1}, \ref{artif2} and \ref{artif3}.\label{dfrac1}}
\end{figure}

In order to verify the validity of equation (\ref{eqconj}), it is necessary to
test our numerical code on artificial 2D signals. Preferably this
should be done on 2D fractals in order to check the efficiency of our method
in finding self-similarity, hence fractality, in the fields of
increments.\\
The basic method for creating artificial fractals is to consider the
spectral density, that is, the mean square fluctuation at any particular
frequency and how that varies with frequency. This technique, the power
spectral method, is based upon the observation that many natural forms and
signals have a $1/\nu^{\beta}$ frequency power spectrum, {\it i.e.} the spectrum falls off
as the inverse of some power of the frequency $\nu$, where the spectral
component $\beta$ is related to the fractal dimension of the signal: $\beta=2H+2$,
with the persistence parameter $0 \le H\le 1$. The latter parameter
quantifies the degree of `roughness' ({\it e.g.} $H=0$ for a fully rough signal;
$H=1$ for a fully smooth signal) and is
related to the fractal dimension, $D$: $H=3-D$ (Moghaddam \etal 1991;
Cox \& Wang 1993). Thus, $D$ will take on values between 2 and 3.\\
Figures \ref{artif1}, \ref{artif2} and \ref{artif3} show three constructions
of artificial fractal signals along with a horizontal cut of the images. To build these signals we first generate
a random 512 $\times$ 512 white-noise signal. We then apply a Fourier transform 
to this image leading to a secondary signal which essentially looks like another
noise field. We scale the resulting spectra by the desired $1/\nu^{\beta}$
function and finally apply an inverse Fourier transform. Controlling how rough
or smooth the surface is has to be done by varying the power relationship.
From Figure \ref{artif1} to Figure \ref{artif3} we have increased
the value of $\beta$, hence decreasing the roughness. Note that
these three images are all based on the same initial seed and therefore have
the same general shape.\\
Figure \ref{dfrac1} (left panels) shows our numerical
applications of equation (\ref{eqconj}) for $p=1$ to the three artificial fractal signals.
Note that choosing $p=1$ in equation (\ref{eqconj}) and the $r^{-2}$ scaling in
equation (\ref{eqdef}) relates any power-law
$\langle|\Delta f(r)| \rangle\sim r^{\zeta(1)}$ to the persistence parameter
$H$: $H=\zeta(1)$. Inspection of Figure \ref{dfrac1} suggests that
the conjecture given in equation (\ref{eqconj}) is very accurate: the derived
slopes are always very close to the theoretical value of $H$.\\
The right panels of Figure \ref{dfrac1} show
the hierarchy of exponents $\zeta(p)$ related to the signals shown in
Figures \ref{artif1}, \ref{artif2} and \ref{artif3}. Here the statistical moments
of $|\Delta f(r;(x,y))|$ have been estimated for each artificial signal, for
$p=$ 1--5, in steps of 0.5. The search for scaling laws has been performed as
before for any individual
value of $p$. Note that our 
artificial signals are monofractal, therefore $\zeta(p)$ should be strictly 
proportional to $p$. This is seen very clearly for $H=0.05$. However,
for $H=0.475$ and $H=0.9$ we detect a clear departure
from this theoretical property for $p>3$ and $p>1$, respectively. This is mainly
due to the small amplitude fluctuations of the last two signals (see Figures
\ref{artif2} and \ref{artif3}, lower panels): for large values of $p$, application of
equation (\ref{eqconj}) may lead to very small numbers whenever
$|T_{\Psi}[f](r;(x,y))|<1$, in combination with dominant extreme
events which are poorly sampled by definition.\\
These numerical experiments
demonstrate how our wavelet method is efficient in detecting any fractal
structure. However, the reliability of the structure function analysis for the highest
orders appears very sensitive to the amplitude of the fluctuations.

\subsubsection{Results on the HST/H$\alpha$ Deep Image of M1-67\label{turbdensity}}

\begin{figure}
\epsfxsize=17cm
\epsfysize=17cm
\epsfbox{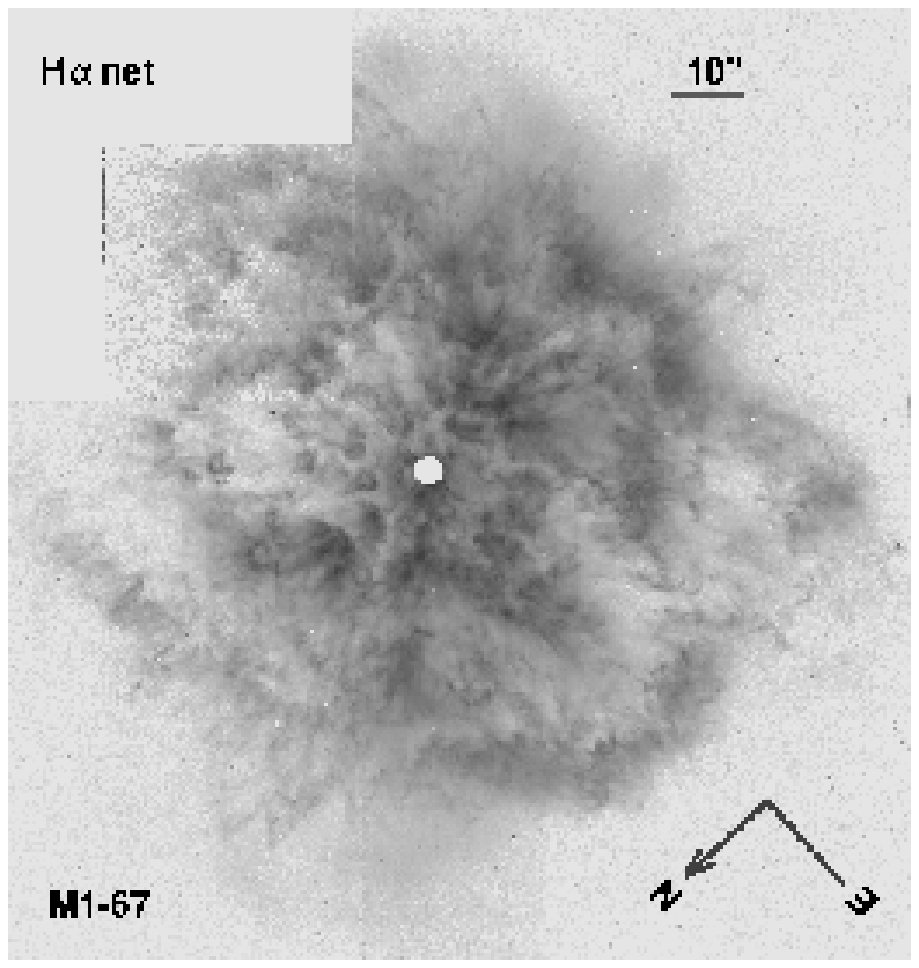}
\caption{Deep WFPC2/H$\alpha$ log-scale image of M1-67 from 4 combined
exposures totalling nearly 3 hours. The field stars have been subtracted
out (enlargement of Figure 1 from Grosdidier \etal 1998).\label{imagem167}}
\end{figure}

\begin{figure}
\epsfxsize=17cm
\epsfysize=17cm
\epsfbox{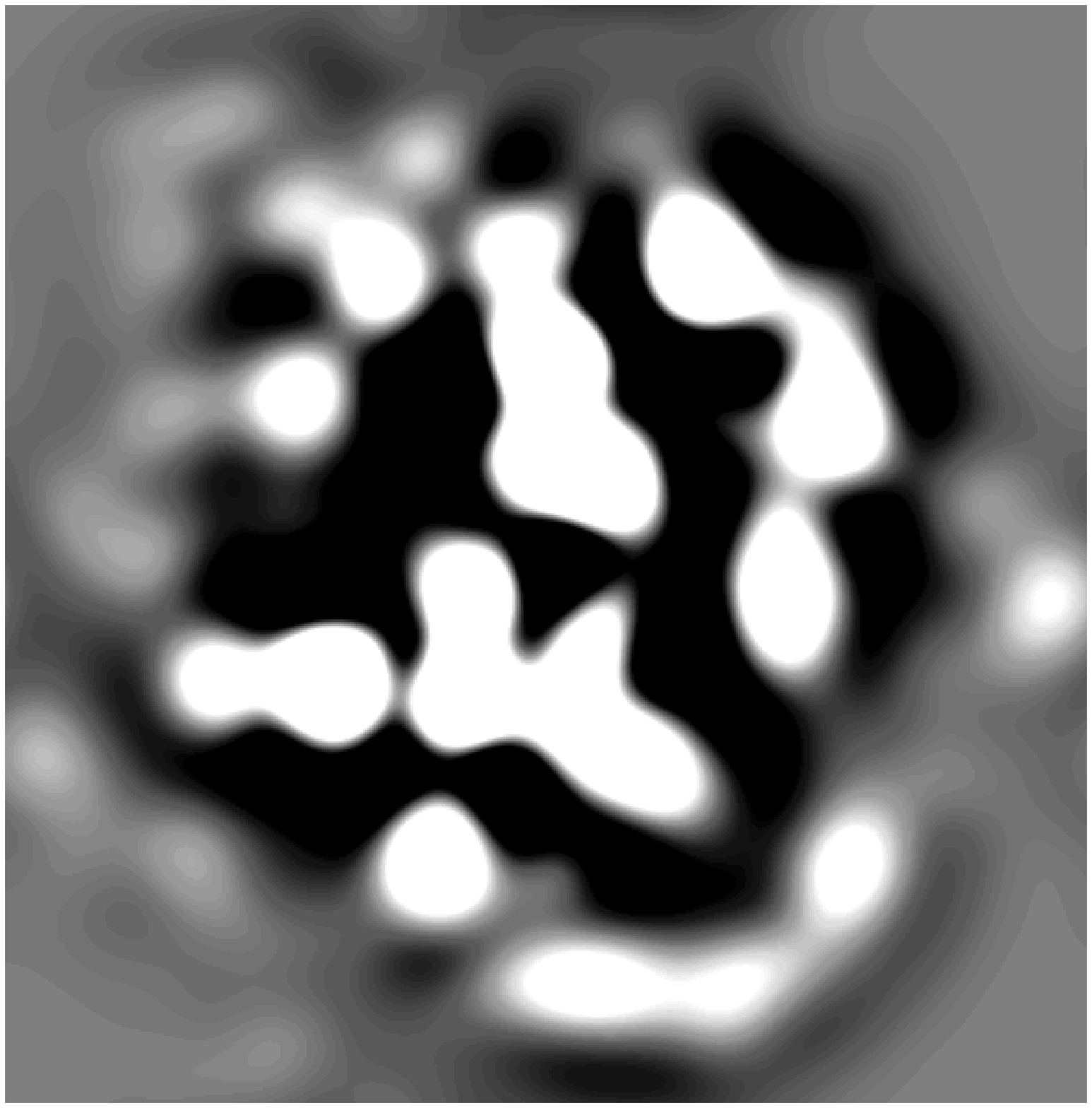}
\caption{Wavelet coefficients of M1-67 at the scale 11.3''. The orientation
of the image is the same as of Figure \ref{imagem167}.\label{aoc1}}
\end{figure}

\begin{figure}
\epsfxsize=17cm
\epsfysize=17cm
\epsfbox{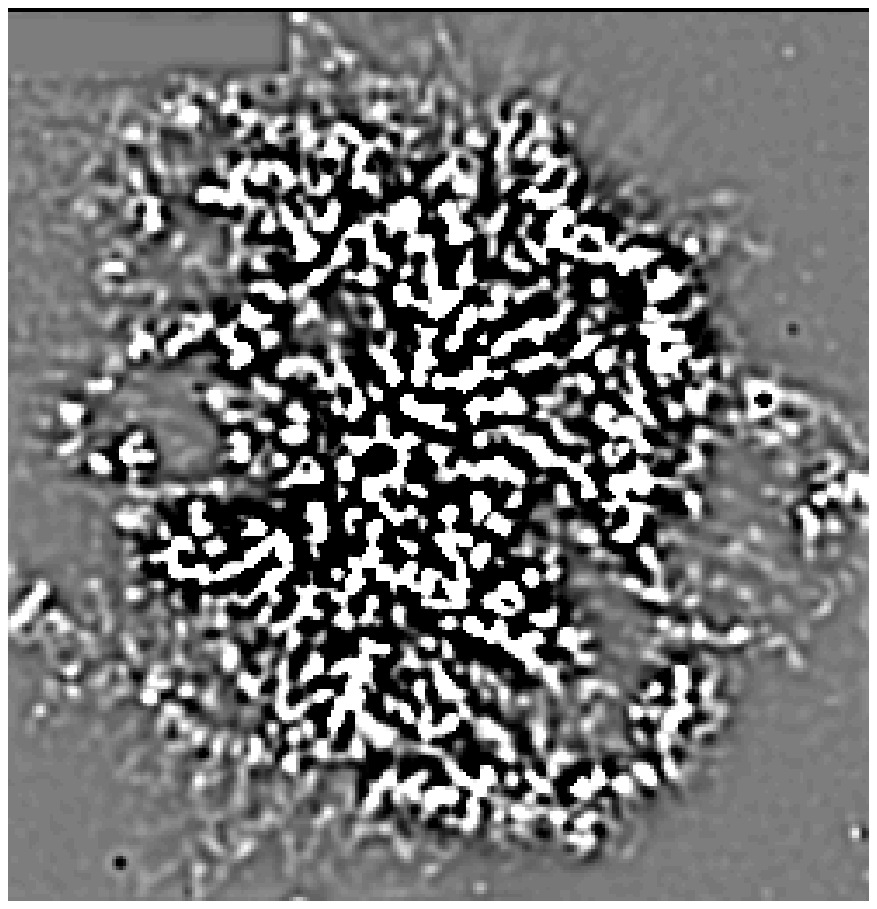}
\caption{Wavelet coefficients of M1-67 at the scale 1.8''. The orientation
of the image is the same as of Figure \ref{imagem167}.\label{aoc3}}
\end{figure}

\begin{figure}
\epsfxsize=17cm
\epsfysize=17cm
\epsfbox{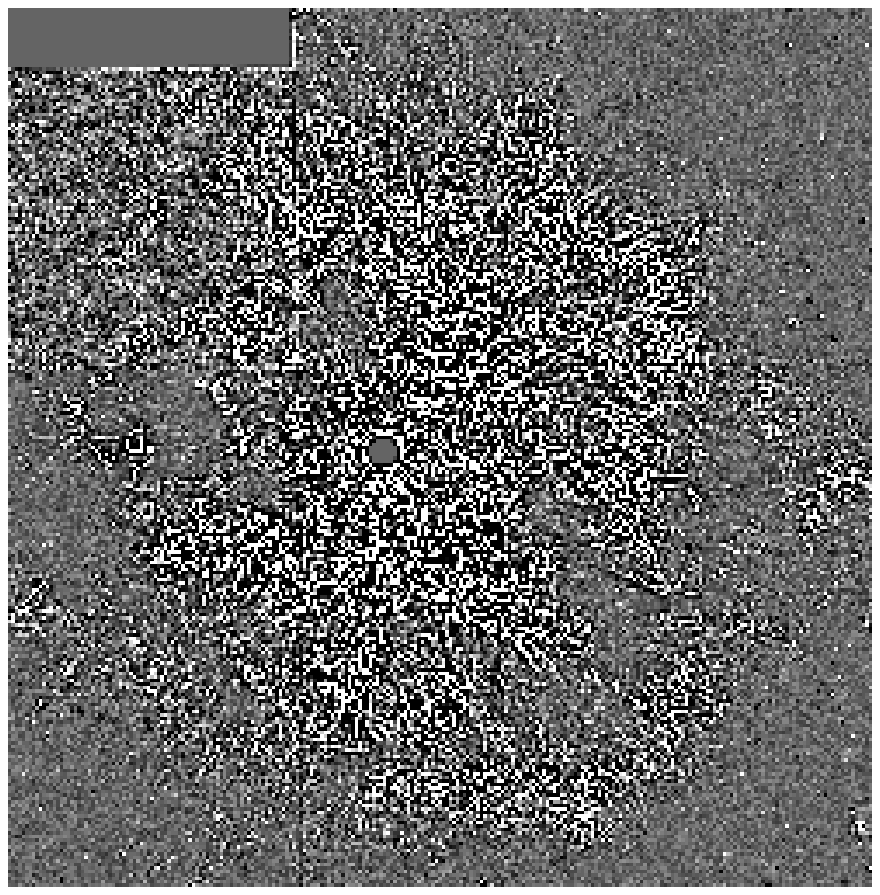}
\caption{Wavelet coefficients of M1-67 at the scale 0.3''. The orientation
of the image is the same as of Figure \ref{imagem167}.\label{aoc2}}
\end{figure}

\begin{figure}
\epsfxsize=17cm
\epsfysize=17cm
\epsfbox{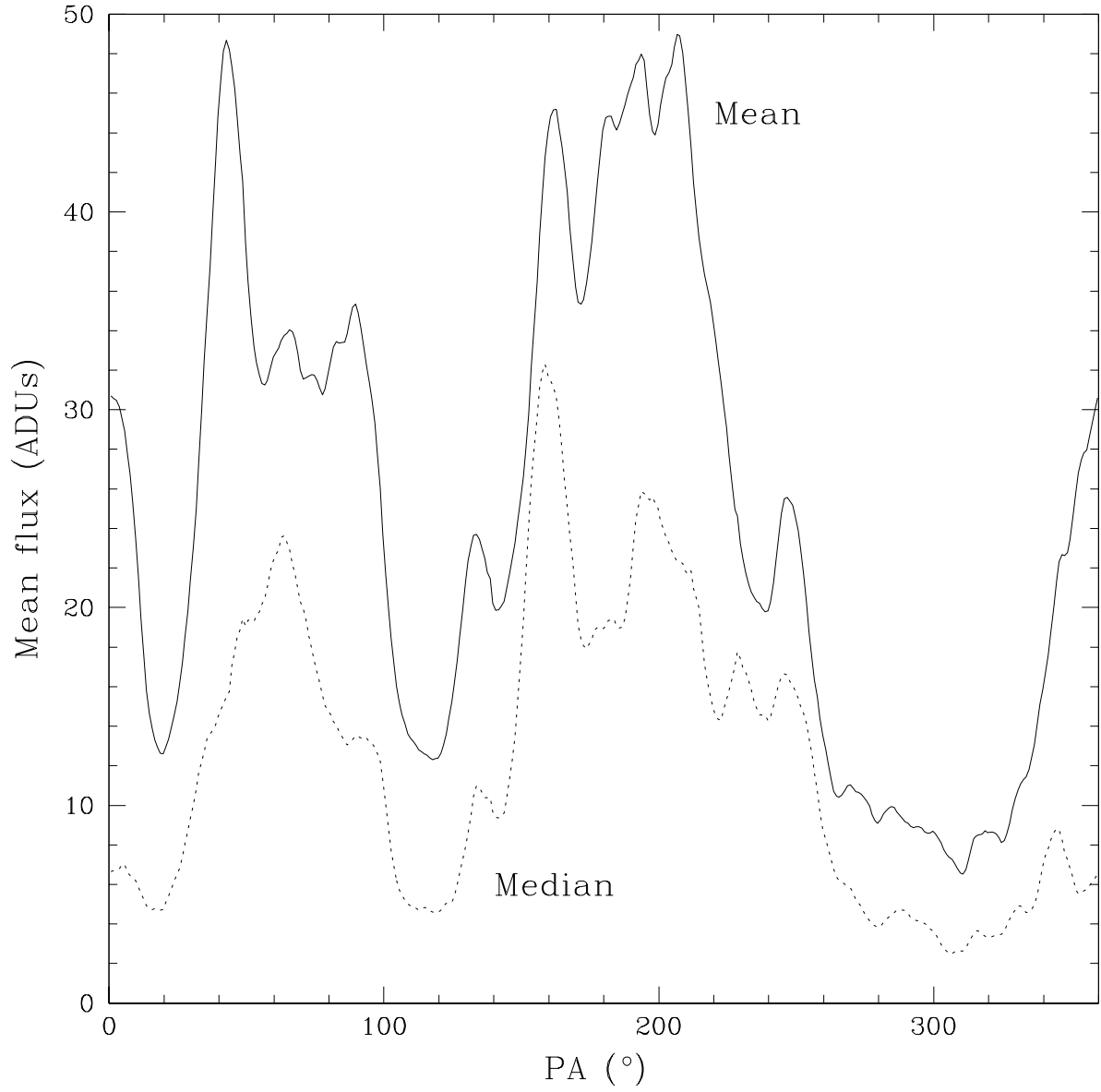}
\caption{The projected angular distribution of the surface brightness
averaged over 10$^{\circ}$-wide sectors centered on WR 124.\label{sector}}
\end{figure}

\begin{figure}
\epsfxsize=14cm
\epsfysize=14cm
\epsfbox{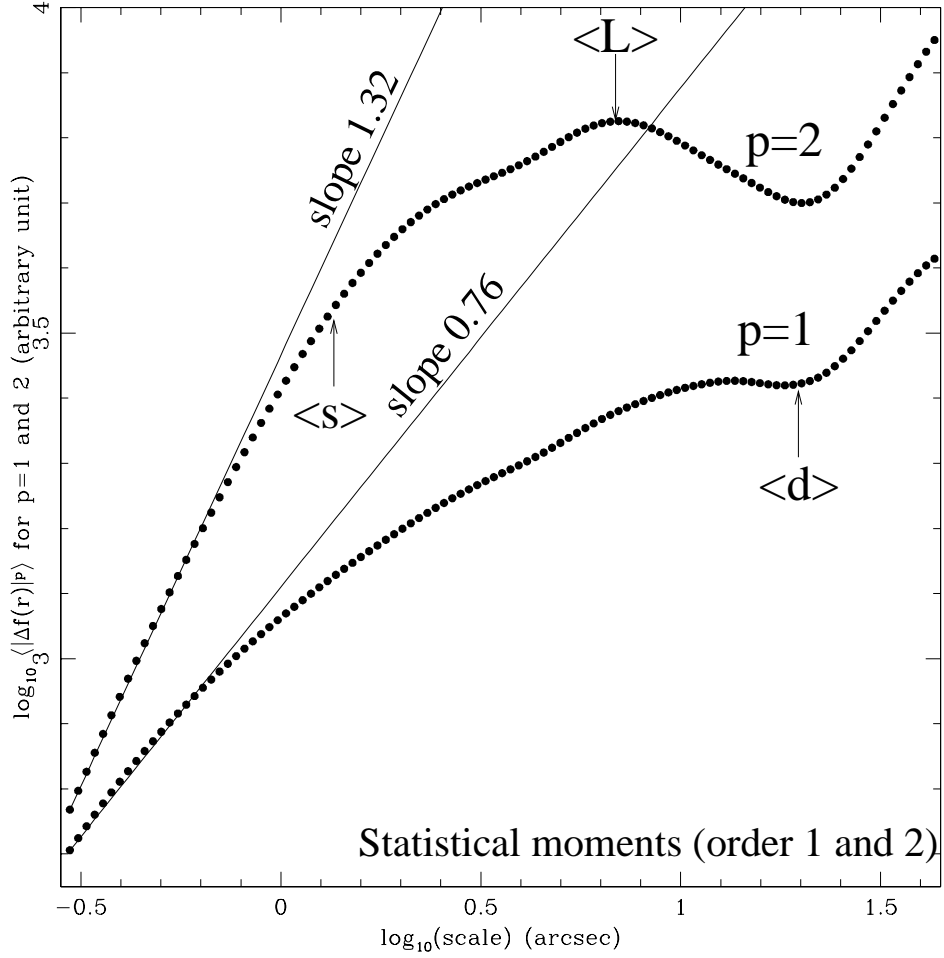}
\caption{Structure function analysis of M1-67. The 
quantities $\langle | \Delta f(r) |^p \rangle$ are plotted down to 3 pixels
of WFPC2 (0.3''), for $p=1$ and $2$. Three characteristic scales are also indicated (see text). One arcsecond corresponds
to about $2.2\times 10^{-2}$ pc assuming a distance of 4.5 kpc for M1-67. 
\label{structure1m167}}
\end{figure}

\begin{figure}
\epsfxsize=14cm
\epsfysize=14cm
\epsfbox{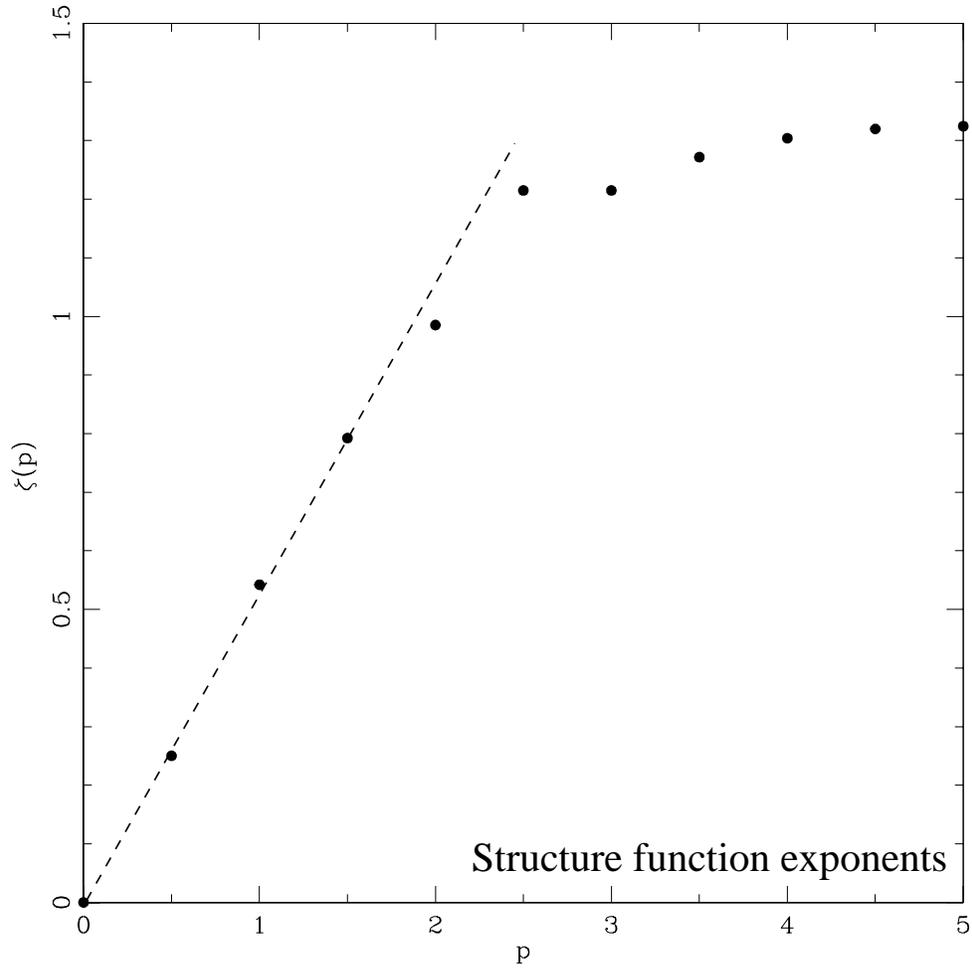}
\caption{Structure function analysis of M1-67. The corresponding $\zeta(p)$ function demonstrates the multi-affinity of
M1-67.\label{structure2m167}}
\end{figure}

Figures \ref{aoc1}, \ref{aoc3} and \ref{aoc2} show, respectively, the wavelet coefficients of the
deep, net H$\alpha$ image (cf. Figure \ref{imagem167}) for three particular
spatial scales. At the largest scale, 11.3'' (Figure \ref{aoc1}), the wavelet coefficients
reveal the large scale structure of M1-67, where one can easily recognize the
bright arcs of Figure \ref{imagem167}. 
The number of clumps then increases rapidly towards smaller scales (see Figures \ref{aoc3} and \ref{aoc2}).
More strikingly, at the smallest scale close to the resolution limit,
0.3'' (Figure \ref{aoc2}), the density field of M1-67 appears remarkably
structured in apparently chaotically oriented filaments {\it everywhere} in the nebula. However, many filaments (especially at the smallest scales) seem to be slightly radially oriented.
Since the field stars have been subtracted off in Figure \ref{imagem167}, these
filaments are strictly related to the fluctuating density field of M1-67.
Note that neither at large scales (Figure \ref{aoc1}), nor at small scales
(Figure \ref{aoc2}) is any bipolar outflow clearly detected. Using the
H$\alpha$ HST image, Figure \ref{sector} shows the
projected angular distribution of the surface brightness averaged over
10$^{\circ}$-wide sectors centered on WR 124. From the median two possible
lobes are detected for $PA\approx 65^{\circ}$, and $PA\approx 185^{\circ}$. 
Since the lobes are not anti-aligned on the plane of the sky, they are unlikely related to a bipolar
outflow. Considering the mean, the situation is even worse: many subpeaks
reveal the highly irregular and patchy structure of M1-67. Using wider sectors (hence losing angular resolution), we arrive at the same conclusions.\\
In Figure \ref{structure1m167} the quantities
$\langle | \Delta f(r) |^p \rangle$ are plotted down to 3 pixels (0.3'') of WFPC2, in order to secure good sampling of the analyzing wavelet, for $p=1$ and $2$. (The contribution of the
instrumental noise has been subtracted off by performing the same analysis on a
rectangular region outside the nebula; the field stars do not contribute
in this calculation since they have been removed.) Note the
short, marginal inertial range
(less than a decade) at the smallest scales from about $6.7\times 10^{-3}$
to about $1.5\times 10^{-2}$ pc which could be
attributed to turbulence. More
strikingly, we detect the presence of a strong scale-break at scales of
$\langle L\rangle\approx$ 5--10'' (or 0.11--0.22 pc, adopting a distance of 4.5 kpc for M1-67),
corresponding to the characteristic widths of the bright arcs seen in
Figure \ref{imagem167}. This suggests that the bright arcs may
be related to a specific earlier {\it ejection} event or hydrodynamic
instability. Looking carefully at Figure \ref{structure1m167} a second
characteristic scale is found at $\langle s\rangle\approx$ 1--2'' (or around $3.3\times 10^{-2}$ pc), where the second-order structure function is still growing but at a significantly lower rate. We interpret these characteristic scales as the presence of essentially two types of
structures in the signal: bright, small $\langle s\rangle$-scale features, and relatively less intense, larger $\langle L\rangle$-scale features. Indeed, as seen in the second-order structure function which emphasizes the most intense events, as the scale $r$ increases $\langle | \Delta f(r) |^2 \rangle$ first increases sharply because the increments from the bottom to the top of the $\langle s\rangle$-scale features dominate and there are more
and more of these. Then $\langle | \Delta f(r) |^2 \rangle$ becomes saturated with respect to these features: for all scales above $\langle s\rangle$, $\langle | \Delta f(r) |^2 \rangle$ contributes a relatively fixed number of events to the spatial average. Further on we see the larger $\langle L\rangle$-scale features at work in a similar way. They allow $\langle | \Delta f(r) |^2 \rangle$ to keep growing but at a slower rate because smaller increments are spawned by these features. For scales larger than $\langle L\rangle$, the second-order structure function again becomes saturated. For scales close to the characteristic projected spatial separation between $\langle s\rangle$- and $\langle L\rangle$-scale features, $\langle | \Delta f(r) |^2 \rangle$ should exhibit a minimum, which is indeed the case for scales around $\langle d\rangle\approx$ 20'', {\it i.e.} 0.4 pc (see Figure \ref{structure1m167} and Marshak \etal 1997 for a similar behavior in another physical context). Typical voids of width about 20'' are compatible with the appearance of M1-67 in Figure \ref{imagem167}.\\
Since the structure function of order 1 scales as
$\langle|\Delta f(r)|\rangle\sim r^{0.76}$ ($H\approx 0.76$) at the smallest
scales, we infer the fractal dimension of M1-67 to be about
$D=3-H\approx$ 2.2--2.3. Thus, M1-67 appears neither as a smooth, nor space-filling
structure. For many different nebular objects, we can find
estimates of their related fractal dimension in the literature. Generally
these fractal dimensions were derived through box-counting techniques
on the perimeter of the nebula, hence leading to $1\le D_p \le 2$, the majority
of the values spanning the range 1.2--1.4 (Bazell \& D\'esert 1988; Blacher
\& Perdang 1990; Falgarone {\it et al.} 1991, and references therein). Experiments have shown that in the case of {\it isotropic}
turbulence/fractality, the fractal dimension of the perimeter $D_p$ and
the fractal dimension $D$ of the two-dimensional projected nebula satisfy
the relation: $D_p=D-1$ (Sreenivasan \& Meneveau 1986; Beech 1992). Therefore, assuming that the fractal
dimension $D_p$ of M1-67 does not vary with the angle of projection
({\it i.e.} assuming isotropy of the turbulent status of the nebula), we infer $D_p($M1-67$)\approx$
1.2--1.3, which is in good agreement with the measurements determined in the above
studies and suggests that this parameter takes on universal values.\\
In Figure \ref{structure2m167} we show the hierarchy of exponents
$\zeta(p)$ as a function of $p$. These exponents have been obtained through
fitting of the statistical moments over 2.2 decades in scale,
including the scale break. Note that $\zeta(p)/p$ is no longer constant when
$p$ exceeds 2--2.5. We interpret this fact as evidence for possible intermittency
in the data, {\it i.e.} sporadic and intense `bursts' of high frequency activity.

\subsection{CFHT Fabry-Perot Data: Structure Function Analysis}

\subsubsection{General Results on the Velocity Field\label{bipo}}

\begin{figure}
\epsfxsize=17cm
\epsfysize=17cm
\epsfbox{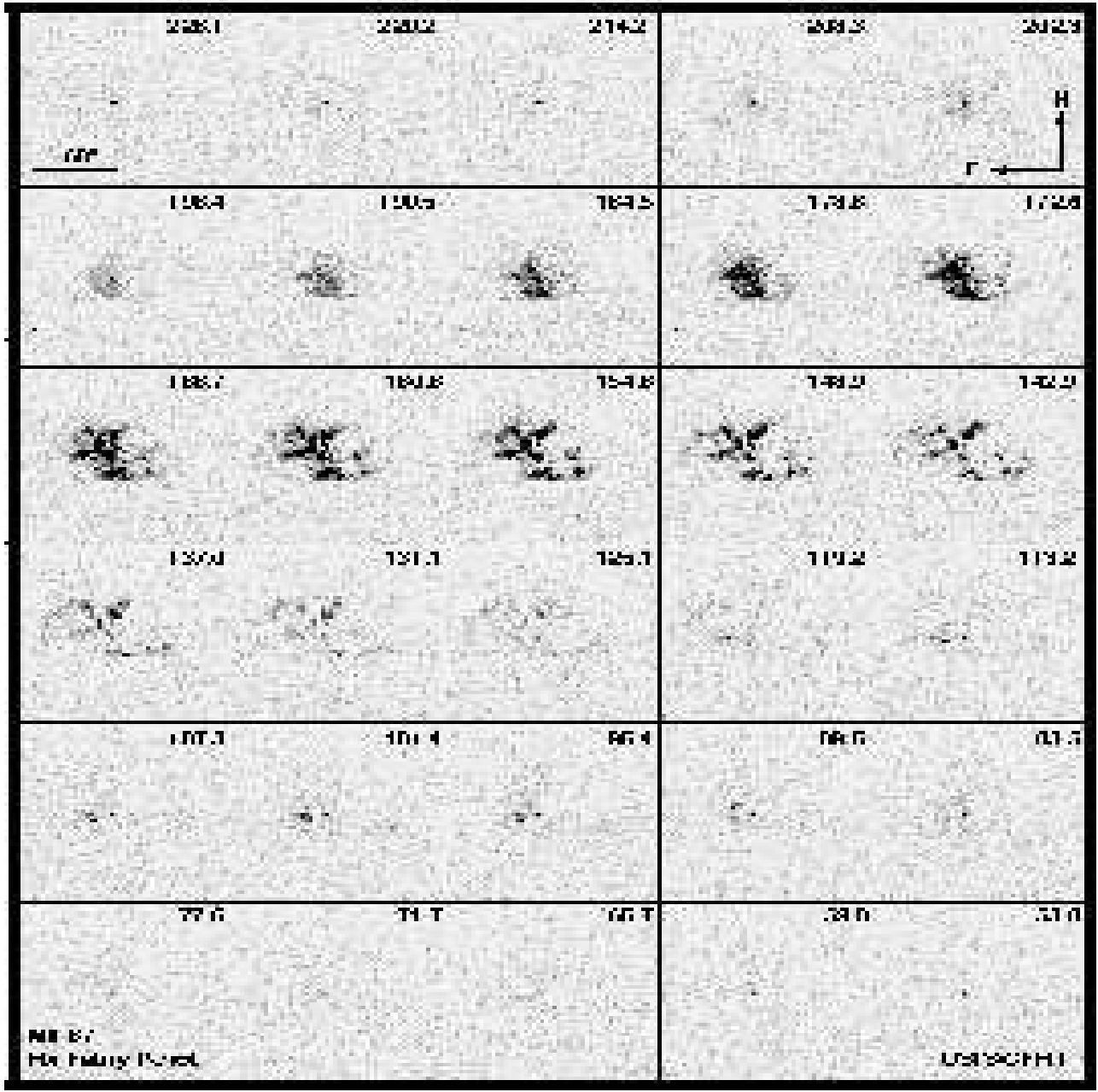}
\caption{Maps of the CFHT/SIS-H$\alpha$ intensity in M1-67 for the
heliocentric radial velocities (in km s$^{-1}$) indicated in the upper
right-hand corners. Everywhere in the nebula except
near the edges, a splitting of the H$\alpha$ line into at least two components
is detected; the high-velocity component is generally brighter
(by a factor 8 or more) than the low-velocity component. The central
star shows up clearly at the same position in each box.\label{velomap}}
\end{figure}

\begin{figure}
\epsfxsize=16cm
\epsfysize=16cm
\epsfbox{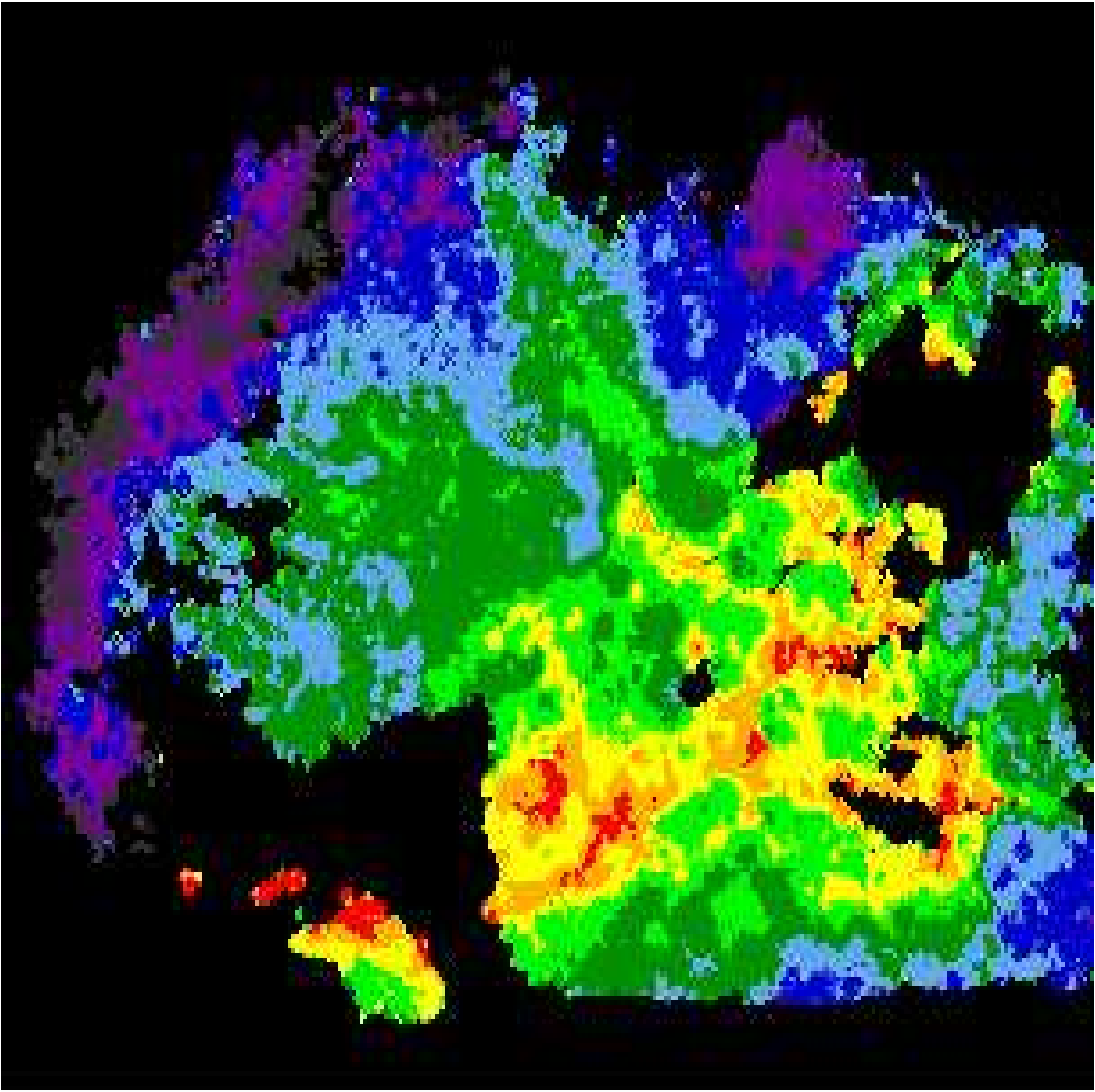}
\caption{Smoothed velocity map of M1-67 for the heliocentric radial velocities of the `red'
component (in km s$^{-1}$). North is up, East to the left.
Note the missing chunk of emission in the
bottom part due to sudden arrival of clouds near the end of the
FP scan.\label{velomap2}}
\end{figure}

\begin{figure}
\epsfxsize=14cm
\epsfysize=14cm
\epsfbox{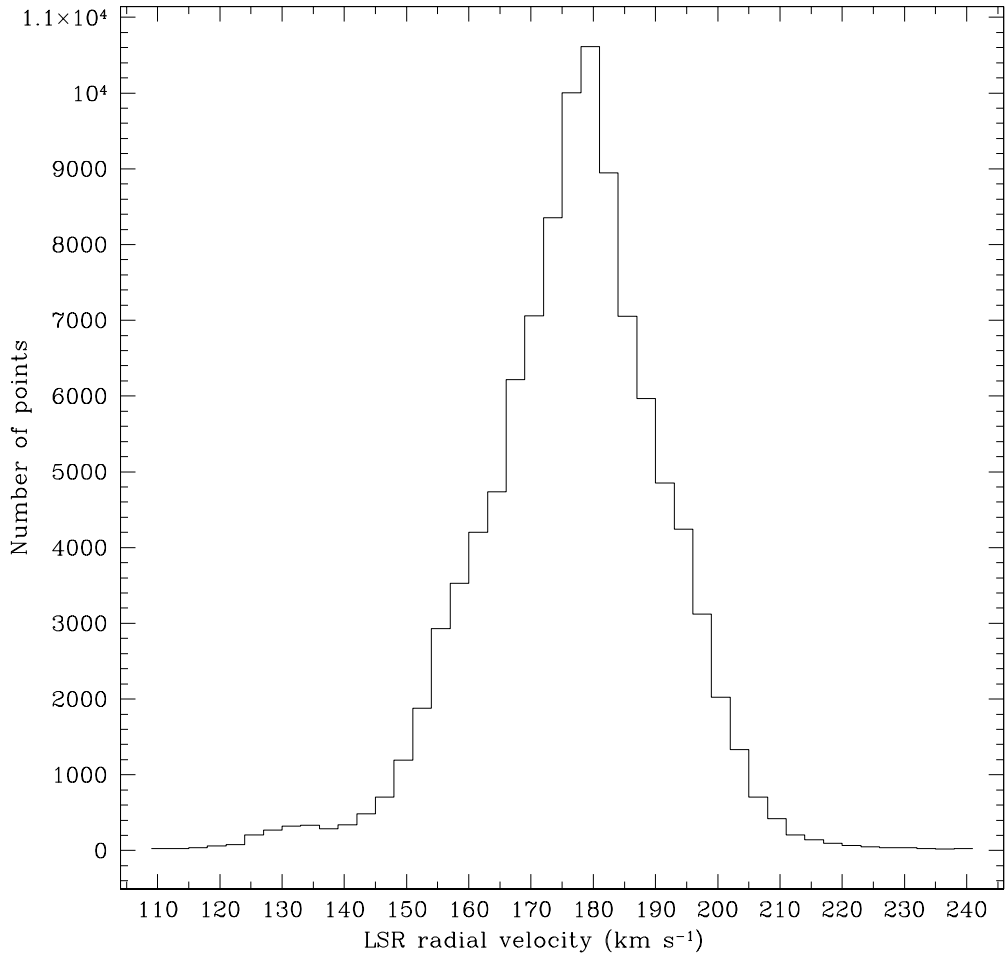}
\caption{The distribution of the 103,287 measured velocity centroids of
the red component.\label{histovelo}}
\end{figure}

Figure \ref{velomap} shows the
channels
covering the spread in radial velocity for the nebula M1-67.
From these data M1-67 appears as a strongly distorted, thick spherical {\it shell} seen
almost exactly along its direction of rapid spatial motion in the ISM
(Moffat \etal 1982; Sirianni \etal 1998). The radial
velocity of the center of expansion is
$\approx$ 137 km s$^{-1}$ (Sirianni \etal 1998). Note that towards the star,
matter is seen at many different velocities. This is due to the broad H$\alpha$
emission line originating in WR124 (wind terminal velocity: $\approx 710$ km
s$^{-1}$; Hamann \etal 1993; Crowther \etal 1995a). Figure \ref{velomap} also shows that the velocity dispersion 
inside M1-67 is quite large: note that a simple thick shell by itself would
not explain the multiple radial velocities along the line of sight. This velocity dispersion leads one to
consider M1-67 as a thick {\it accelerating} shell.
On the whole, instead of appearing as a nice hollow-type shell projected on the
sky, we
probably see the cap of the bowshock nearly straight on from behind,
the far side being greatly intensity-enhanced (by a factor of 8 or more) compared
to the near
side, probably as a result of raming with the ISM.
This was already claimed by Solf \& Carsenty (1982).\\
Despite the presence of a `protective' wind-blown cavity of radius
several 10 pc, expected to have been produced by the original O-star
progenitor, the current nebula of radius $\sim1$ pc will nevertheless be able
to ram the ISM directly.  This is because of the rapid motion of the
central star (c. 200 km s$^{-1}\approx 2\times 10^{-4}$ pc yr$^{-1}$) combined with the braking of
the cavity expansion in the ISM, allowing the nebula to reach the edge
of the cavity and enter in contact with the ISM.\\
The processing of all the
interferograms enabled the measurement of the velocity centroids covering most of the
surface of the nebula. Because of the low signal-to-noise ratio of the approaching,
near side of M1-67, this has been done only for the far, `red' side of the nebula,
{\it i.e.} for velocities greater than 137 km s$^{-1}$.
Figure \ref{velomap2} shows the velocity map of M1-67 for the red component.
From this figure, M1-67 exhibits a distorted velocity field reminiscent of a thick
expanding shell as suggested by Figure \ref{velomap}. In addition, the velocity
field does not show any clear bipolar outflow, contrary to the previous claim
by  Sirianni \etal (1998). We argue that the Sirianni \etal conclusions were
based on an insufficient amount of data due to poor spatial sampling of the
velocity field. Note that sudden, huge changes in the transparency
conditions of the sky occurred during the scanning at the end of the night.
This explains the absence of measurements in the extreme south part of
M1-67.\\
Figure \ref{histovelo} shows the distribution of the 103,287 measured velocity points of
the red component. The mean LSR velocity is 176.2 km s$^{-1}$ with a
standard deviation of 15.3 km s$^{-1}$. 
Given the systemic heliocentric radial motion of M1-67 ($\approx 137$ km s$^{-1}$)
and assuming a symmetric expansion of the nebula,
with a velocity of expansion of about 46 km s$^{-1}$ (Sirianni \etal 1998),
the maximum expected radial velocity should be $\approx$ 183 km s$^{-1}$. 
However, we note numerous points above this value (up to about 225 km s$^{-1}$)
showing the high velocity dispersion within M1-67.
Note that Figure \ref{histovelo} shows velocities well below 137 km
s$^{-1}$, down to about 120 km s$^{-1}$, {\it i.e.} velocities
apparently related to
the blue component. This is mainly due to velocity perturbations of the shell
near its periphery, and the resulting difficulty in splitting the two components. 
Moreover, note
the presence of a small bump near 120--140 km s$^{-1}$. Since this value is very close
to the radial velocity of the center of expansion (137 km s$^{-1}$), here we are
mainly concerned
with regions of M1-67 located near the periphery of the nebula. Such a bump
is evidence for velocity perturbations likely due to the interaction with
the interstellar medium. On the whole, the irregular nature of the velocity field is possibly related to either large variations in the density distribution of the ambient ISM,
or large variations in the central star mass-loss history.

\subsubsection{Structure Function Analysis of the Velocity Field}

\begin{figure}
\epsfxsize=16cm
\epsfysize=16cm
\epsfbox{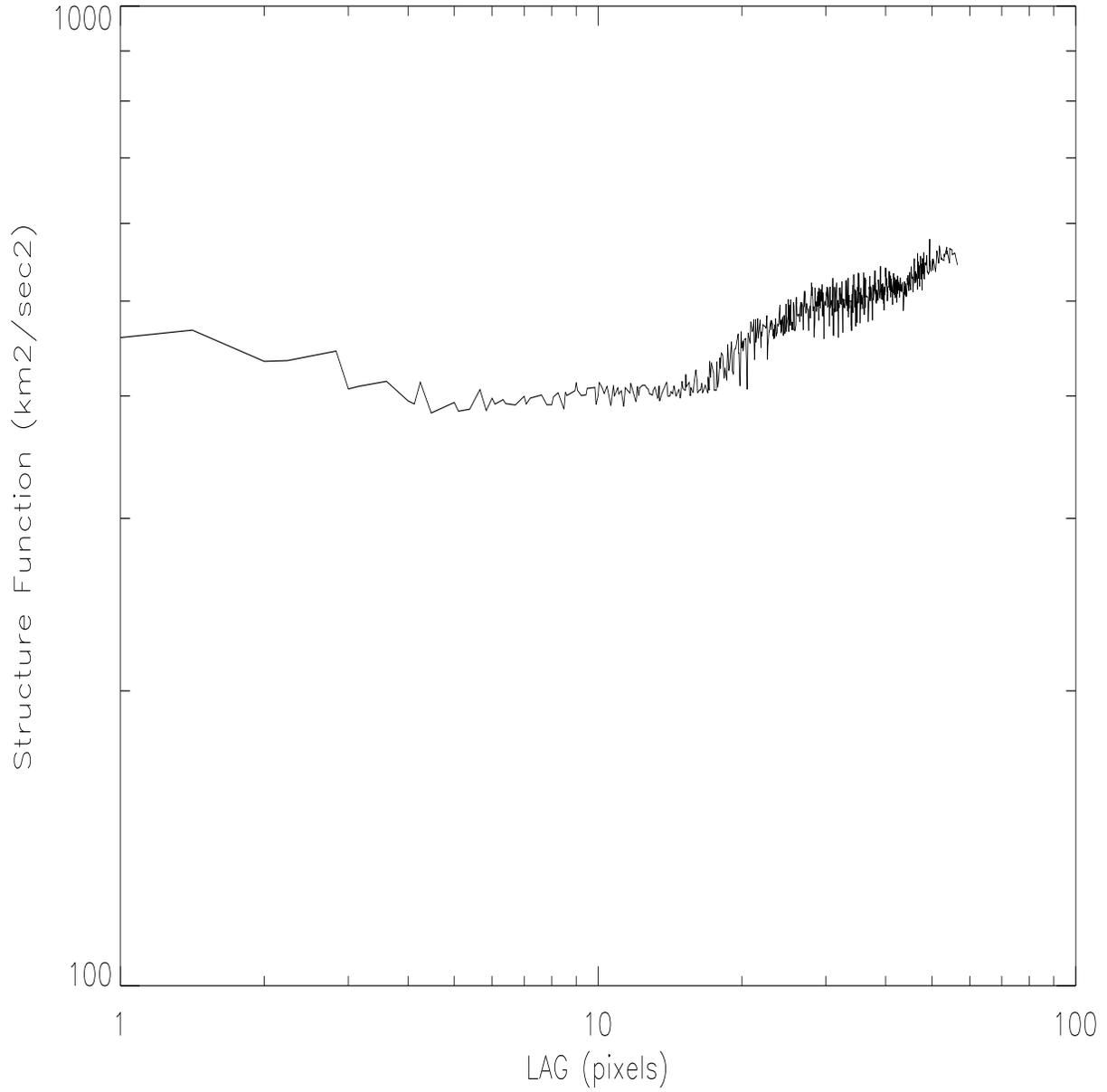}
\caption{The second-order velocity structure function of M1-67 where LAG is the
separation between velocity pairs. One pixel corresponds to
$\approx 6.5\times 10^{-3}$ pc assuming a distance of 4.5 kpc for M1-67.
\label{structurem167velo}}
\end{figure}

Because any valid statistical analysis of velocity fluctuations has to be done
on a (at least local) stationary velocity field, one has to subtract global,
systematic, nonstationary trends from it. Rather than fitting a two-dimensional polynomial
to the velocity data in order to remove the expansion pattern of the distorted shell
(which would generally lead to additional new systematic trends), we
have convolved the centroid velocity map with Zurflueh filters (Zurflueh 1967; Miesch
\& Bally 1994) of different frequency response widths, and removed the smoothed map
from the original data. The Zurflueh filter is a symmetric, low-pass filter
which nicely approximates a step function in the frequency domain and 
moderates the introduction of artificial high-frequency patterns in the convolved
data. This systematic trend correction gave us the residual fluctuating velocity
component on which we have performed the structure function analysis.
Figure \ref{structurem167velo} shows the computed second-order discrete structure
function of the velocity field of M1-67:

\begin{equation}\label{defstructure}
S(r)=\frac{\Sigma \left[ v(X)-v(X+r)\right]^2}{N(r)},
\end{equation}

where $X$ and $r$ are two-dimensional vectors on the plane of the sky, the
summation being done on all the pairs separated by $r$, $N(r)$ (Scalo 1984;
Miesch \& Bally 1994).
In Figure \ref{structurem167velo}, the second-order velocity structure function
shows no clear, inertial range, as well at small as at larger spatial
separations. The absence of a well-defined inertial range appears very likely
because it is independent of the width of the applied Zurflueh filter.
However, with a distance of 4.5 kpc to M1-67 and a pixel size of 0.3'', we
detect a possible, marginally inertial regime occuring in the short range 16--50
pixels, {\it i.e.} 0.1--0.3 pc. A linear least-squares' fit of the slope in
the range where there is a possible correlation of the results has been done
using a power-law and gives a slope of $\zeta(2)\approx0.3$ ({\it i.e.} the related
distribution of turbulent kinetic energy with spatial scale would be described
by an energy spectrum, $E(k)$, scaling as $k^{-1.3}$ with the wave number $k$).
Such a value does not meet the Kolmogorov type description for which one would
obtain $\zeta(2)=2/3$ ({\it i.e.} $E(k)\sim k^{-5/3}$). On the other hand, the 16--50 pixel possible
inertial
range is so short that it is unlikely significant or real.\\
At smaller scales, down to the achieved spatial resolution, the 2--16
pixel range ({\it i.e.} 0.01--0.1 pc) seems to be related with an almost
stationary, scale-independent field of increments. Indeed, in
this range $S(r)$ is found to scale trivially: $\zeta(2)\approx 0$.
Note we are not able
to confirm the inertial range detected from the density field (see section
\ref{turbdensity}) because this inertial range ends at about 0.3'', which is
well below the resolution of the CFHT data. Similar results are obtained when estimating the second-order structure function of the centroid velocity field using (linear-trends free) wavelet coefficients following the method described in section \ref{method}.

\subsubsection{The Distribution of Velocity Increments in M1-67 --- Structure Function Analysis of the Regularized Velocity Map}

\begin{figure}
\epsfxsize=14cm
\epsfysize=14cm
\epsfbox{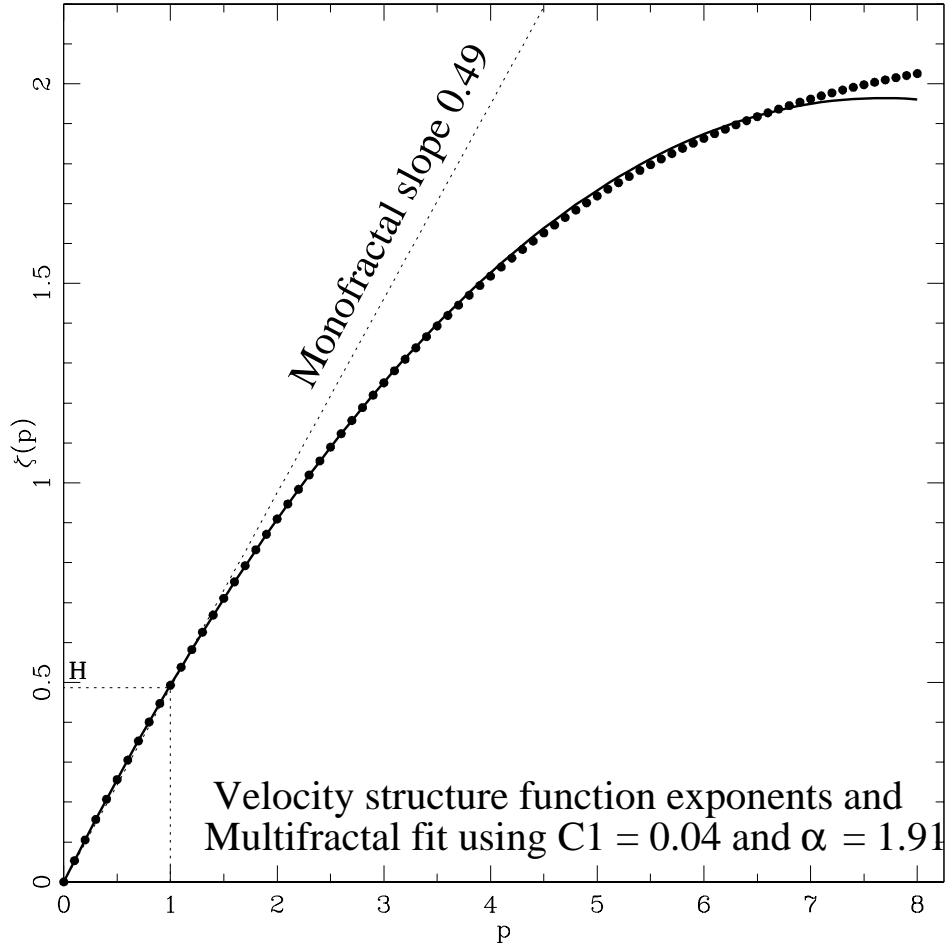}
\caption{Velocity structure function analysis of M1-67 after small-scale regularization using a 1-pixel radius ring median filter. The corresponding $\zeta(p)$ function (points) demonstrates the multi-affinity of the velocity field of
M1-67. A universal multifractal fit is also shown as a solid curve (see text).\label{vstructure2m167}}
\end{figure}

\begin{figure}
\epsfxsize=14cm
\epsfysize=14cm
\epsfbox{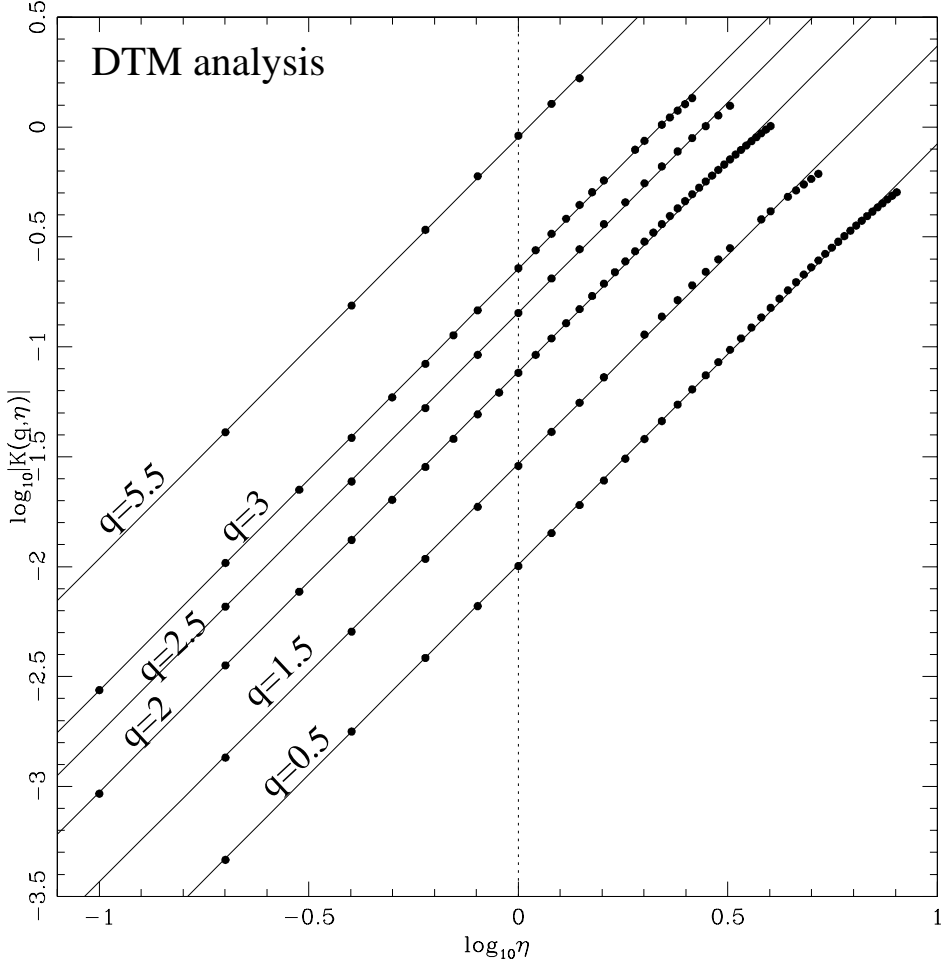}
\caption{Plots of $\log_{10}|K(q,\eta)|$ (using $q=0.5$, 1.5, 2, 2.5, 3 and 5.5) obtained with a DTM analysis of the velocity field of M1-67
after small-scale regularization. From the slopes and the $\log_{10}\eta=0$ intercepts of the linear region we obtain 
$\alpha=1.90$--1.92, and $C_1=0.04\pm0.01$. There is a first order multifractal phase transition near $\max(q\eta,\eta)=7$,
which is a consequence of the finiteness of the analyzed sample.\label{singulm167}}
\end{figure}

Because we depend on the spatial coordinates to perform all averaging
operations when analysing the velocity data, our first task is to determine
which aspects of the signal are more likely to be stationary.
The velocity field of M1-67 is non-stationary but with presumably
locally stationary increments, so that in the previous section we turned to structure
function analysis.
Although possibly related to effects of finite spatial resolution (see
Marshak \etal 1994, and references therein), the small, almost null (at
the smallest scales), value of $\zeta(2)$ derived from the velocity field
suggests that the field of increments $\Delta v(X;r)=v(X)-v(X+r)$ in equation
(\ref{defstructure}) is {\it apparently} essentially scale-independent.\\
One-point probability density functions (PDF) make no use of the ordering of
the data, leaving wide open the possibility of correlations between
data-values at different points or structures in the data. However, the
apparent scale independence of the velocity increments relaxes this
drawback of one-point PDF statistics. One traditional approach is one-point
PDF histograms, typically searching for gaussianity.\\
Figure \ref{levy} shows the empirical PDFs of velocity increments as traced by the velocity field wavelet coefficients for four different spatial separations.
Each PDF is compared with a L\'evy stable distribution (found using the computer program STABLE; Nolan 1997) and the best-fit Gaussian distribution. Recall that L\'evy stable distributions arise from the generalization of the central limit theorem to a wider class of distributions. In the Gaussian case, the renormalized sum of independent and identically distributed random variables with finite variance has the same probability distribution as the individual terms in the sum. In 1937 L\'evy showed that the Gaussian distribution is but a special case in a wider family of distributions once the constraint of finite variance is relaxed (Feller 1971). The resulting distributions are referred to as L\'evy stable distributions, or simply L\'evy distributions, and are characterized by slowly decaying tails and infinite moments. L\'evy distributions, $L_{\alpha}$, require four parameters to describe: an index of stability $\alpha$ in the interval (0,2], a skewness parameter $-1\le\beta\le 1$, a scale parameter $\sigma>0$ and a location parameter $\mu$. The index $\alpha$ determines the rate at which the tails of the distribution decrease ($L_{\alpha}(x)\propto x^{-1-\alpha}$); when $\alpha=2$, a Gaussian distribution results. When $\alpha<2$ the variance is infinite. In general, the $p$-th moment of a L\'evy stable random variable is finite if and only if $p<\alpha$. 
The parameters $\sigma$ and $\mu$ determine the width and the shift of the peak of the distribution.
For a positive (resp. negative) $\beta$ the distribution is skewed to the right (resp. the left). For $\beta=0$, the distribution is symmetric about $\mu$. In addition, as $\alpha$ approaches 2, $\beta$ loses its effect and the distribution approaches a Gaussian regardless of $\beta$ (Nolan 1997).\\
Strictly speaking, velocity increments cannot have a L\'evy stable distribution because this would imply nonzero probability density in regions outside the range for which velocity increments are physically meaningful. However, these distributions can yield useful approximations over a truncated/finite range as demonstrated in other physical contexts (Painter \etal 1995). In addition, when we use an infinite variance distribution in order to fit the distribution of velocity increments, all that we are saying is that the observed slowly decaying tails result in behavior similar to that of the infinite variance model.  On the whole, we stress the fact that variance is but one measure of spread for a distribution, and is not appropriate for all problems (Painter \etal 1995).\\
As the scale increases, Figure \ref{levy} shows us that the distribution approaches a Gaussian distribution, as expected for a truncated L\'evy distribution (Nakao 2000). At small scales, the distribution of the velocity increments is characterized by a non-negligible number of extreme events making the PDF of the velocity increments not narrow enough to enable a simple dimensional argument to relate quantitatively all moments: $\langle|\Delta v(r)|^p\rangle$ no longer approximates $\langle|\Delta v(r)|\rangle^p$ and precludes the determination of any actual (multi-)fractal structure. At large scales the gaussianity of the PDFs releases this drawback and non-trivial scaling laws may be detected, although likely biased, as seen in Figure \ref{structurem167velo}. Note that the L\'evy fits in Figure \ref{levy} are nearly symmetric ($\beta\approx 0$) at small scales. However, as the scale increases, $\beta$ increases mainly because the wavelet coefficients take into account more and more non-linear trends in the signal.\\
Because of the high spatial resolution of our Fabry-Perot observations and
the good seeing conditions, one expects each pixel to look at a fairly
independent piece of the sky. However, very
low flux regions could see their spectrum polluted by the psf wings of
a very bright neighbor. Such combination is indeed found in our data and leads to the dominant, larger velocity increments, in line with the $\langle s\rangle$-scale intense features detected in the density field wavelet analysis (see section \ref{turbdensity}). As a consequence, the velocity centroid measurements are sparsely corrupted resulting in fat-tail PDFs of velocity increments.
Note that, although
there is a regular grid of pixels, the emission is distributed
irregularly according to the density distribution. Should the
velocity signal be coming primarily from the brightest regions
then the scaling signal could be additionally confused by the irregular grid
signal (see Rauzy \etal 1995). However, our FP, high throughput velocity map is not found to
suffer from a significant interference signal from a non-uniform grid.\\
We therefore applied a 1-pixel radius ring median filter to the velocity map. The ring median filter is defined as a median filter which assigns weights only to selected pixels in an annulus. Its advantage is that it has a sharply-defined scale
length; that is, all objects with a scale-size less than the radius of the ring are filtered and replaced by the local background level. It provides a
simple method to remove noise-like deviations (independent of morphology) from a digital image, leaving behind the large-scale structures
and overall gradients (Secker 1995). We expect each velocity point of the regularized velocity map to be an accurate measure of the underlying
gas kinematics, averaged on its pixel seeing disk.\\
Figure \ref{vstructure1m167} shows the computed structure
functions of the velocity field of M1-67 after small-scale regularization  using a 1-pixel radius ring median filter. The scaling range [$\approx 1.4$'',10''] where the exponents are defined is clearly marked. The break which defines the upper limit of $\approx$ 10'' is not robust with respect to adding to the average more data points, and simply reflects a violation of the ergodicity assumption. Below the 1.4'' scale, the slope is slightly steeper, as a consequence of the ring median filtering which tends to smooth the signal at the smallest scales. Indeed, using a 2-pixel ring annulus simply results in moving the lower limit for scaling towards larger scales. Therefore, the cleaned velocity map is found to scale over {\it at least} one order of magnitude in scale, with $\zeta(1)\approx 0.48$--0.49 and $\zeta(2)\approx 0.90$--0.91. The latter value is close to
1.0 ($E(k)\sim k^{-1.9}$), suggesting the dynamics is essentially shock dominated (for which $E(k)\sim k^{-2}$). Figure \ref{vstructure2m167} shows the function $\zeta$ as a function of $p$, up to 8.0, in steps of 0.1. Clearly $\zeta(p)/p$ is non-linear and suggests the multi-affinity of the velocity field.\\
Multi-affine modelling could be done in the context of `Universal Multifractals' (Schertzer \& Lovejoy 1987) where multifractals are the generic result of multiplicative cascades. A continuous-scale limit of such
processes leads to the family of log-infinitely divisible distributions which have a Gaussian or L\'evy stable generator (Schertzer \& Lovejoy 1987, 1997; Lovejoy \& Schertzer 1990). For universal multifractals we have:
\begin{eqnarray}\label{multiuniv}
\zeta(p)=p\zeta(1)-\frac{C_1}{\alpha - 1}(p^{\alpha} - p) & (\alpha \ne 1), \\
\zeta(p)=p\zeta(1)-C_1 p \ln (p) & (\alpha = 1),
\end{eqnarray}

where $C_1 \le 2$ is an intermittency parameter and $0< \alpha \le 2$ is the L\'evy tail index. After the existence of a scaling regime is established, one can test for universal multifractal behavior. An efficient technique for that purpose is the `Double Trace Moment' (DTM) method (Lavall\'ee 1991). A new function $K(q,\eta)$ is first defined as $K(q\eta)-qK(\eta)$, with $K(q)=q\zeta(1)-\zeta(q)$. In the case of universal multifractals, the $\eta$-dependence in $K(q,\eta)$ factorizes as: $K(q,\eta)=\eta^{\alpha}K(q)$. The DTM method allows the computation of $K(q,\eta)$, and assuming universality, the L\'evy index $\alpha$ can be estimated by fixing $q$ and varying $\eta$. With a good estimate of $\alpha$ in hand, an ordinary least-squares estimate of $C_1$ is easy to obtain. In our case, a standard two-parameter nonlinear regression does very fine over almost two orders of magnitude in $\eta$, with $\alpha\approx 1.91$ and $C_1\approx 0.04 \pm 0.01$ (see Figure \ref{singulm167}). The related $\zeta(p)$ universal multifractal fit is shown in Figure \ref{vstructure2m167} (solid curve). The agreement between the empirical $\zeta(p)$ and the multifractal fit is very good for $p$ below $\approx$ 7. Note that the finite size of the sample implies that sufficiently high order moments are dominated by the largest values in the field and therefore underestimate the true ensemble moments. Beyond $\max(q\eta,\eta)=(2/C_1)^{1/\alpha}\approx 7.7$ equation (\ref{multiuniv}) is no longer true, and $\zeta(p)$ becomes linear (see Figure \ref{vstructure2m167}). This is the so-called first order multifractal transition (Schertzer \& Lovejoy 1992) which is also seen in Figure \ref{singulm167} where the empirical $K(q,\eta)$'s are no longer fitted by the linear regressions.

\section{Discussion and Conclusion}
\begin{figure}
\epsfxsize=16cm
\epsfysize=16cm
\epsfbox{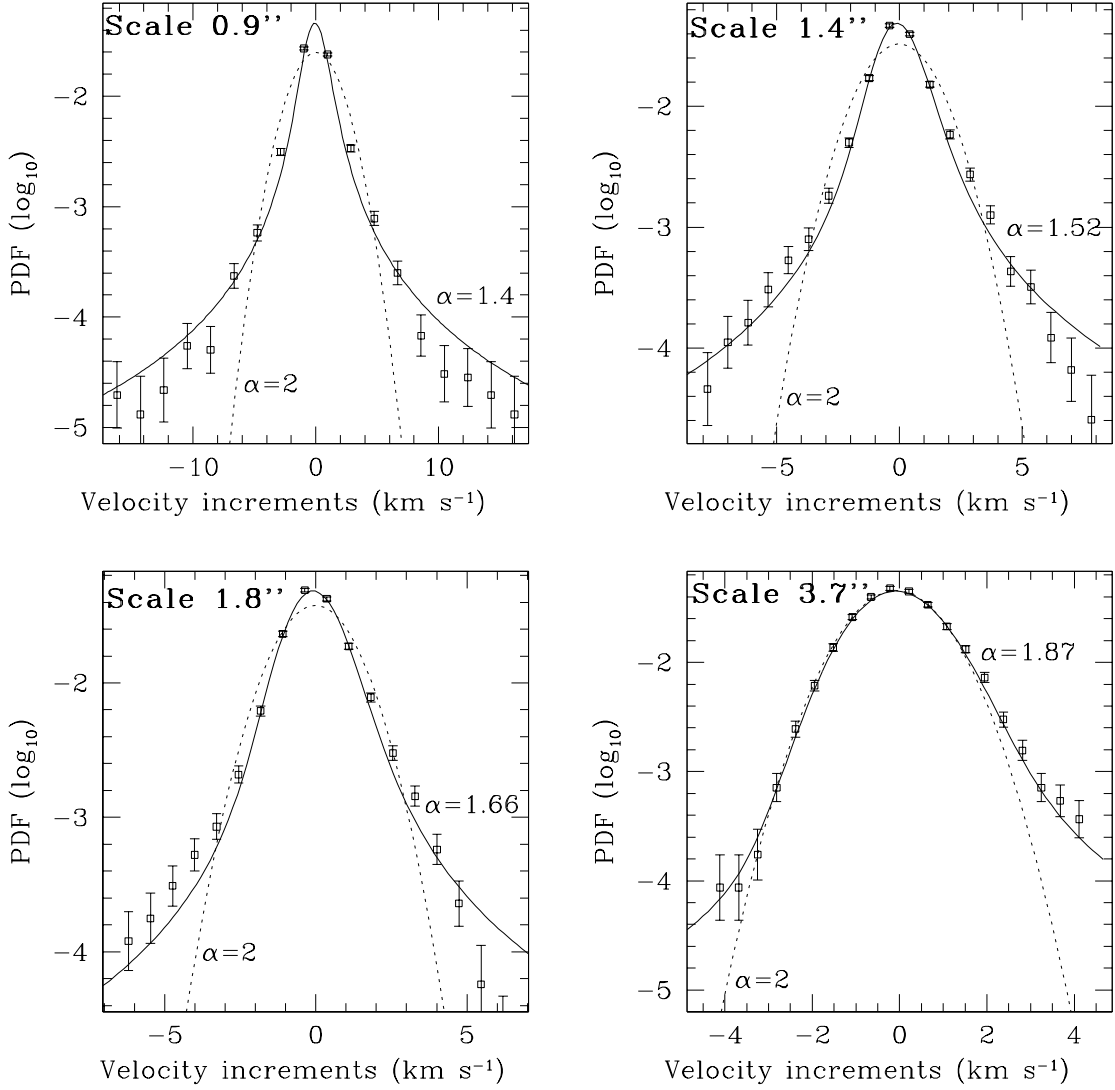}
\caption{Empirical PDFs of velocity increments as traced by the velocity
field wavelet coefficients for four different spatial scale separations (squares). Each PDF is
compared with a L\'evy stable distribution (solid curve) and the best-fit Gaussian
distribution (dotted curve). The histograms were constructed from about 27300 individual
measurements at each scale.\label{levy}}
\end{figure}

\begin{figure}
\epsfxsize=14cm
\epsfysize=14cm
\epsfbox{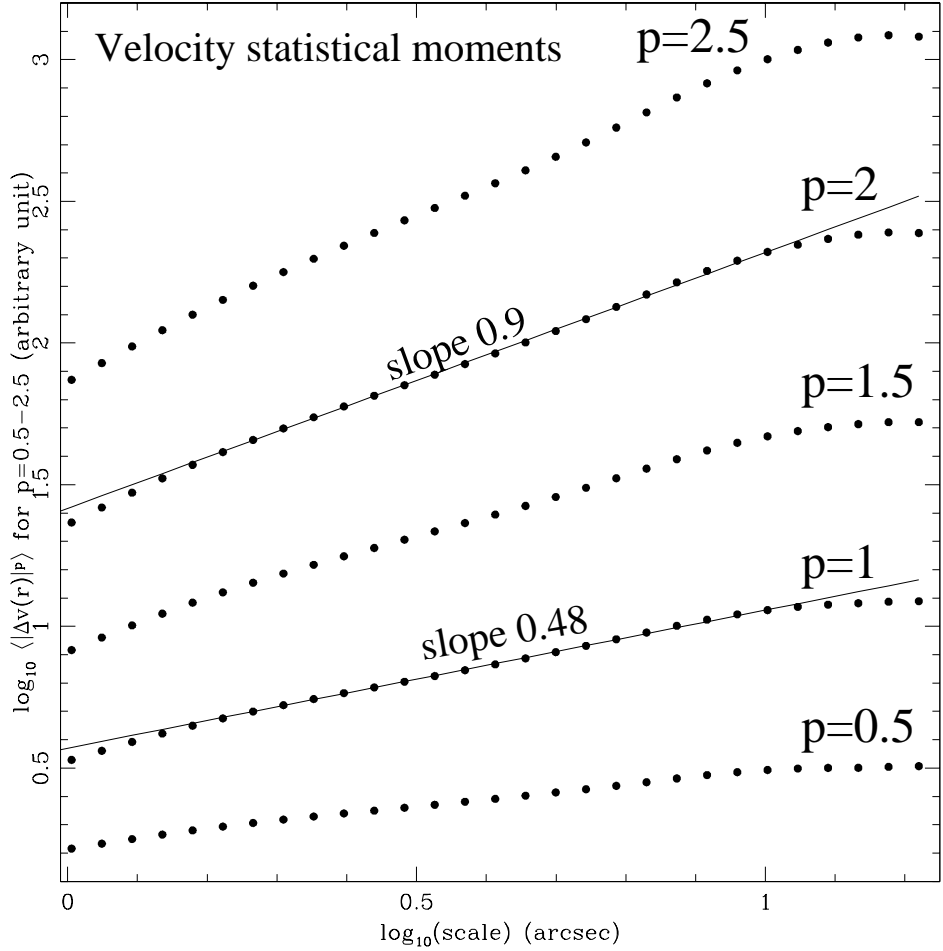}
\caption{Velocity structure function analysis of M1-67 after small-scale regularization using a 1-pixel radius ring median filter. The 
quantities $\langle | \Delta v(r) |^p \rangle$ are plotted down to 3 pixels
of CFHT/SIS (0.9''), for $p$ = 0.5--2.5, in steps of 0.5. One arcsecond corresponds
to about $2.2\times 10^{-2}$ pc assuming a distance of 4.5 kpc for M1-67.
\label{vstructure1m167}}
\end{figure}

Using the unique, high-resolution, wide-field imaging capabilities
of HST, we have tested quantitatively, via scaling laws, for compressible
turbulence in the distribution of density clumps in the ejected nebula M1-67.
These data have been straddled by complementary CFHT Fabry-Perot H$\alpha$ data
in order to perform the same test on the velocity field of M1-67.
What is generally understood as turbulence is the difference resulting
from the subtraction of the instrumental and thermal broadening from the
observed emission line widths. In other words, turbulence is a quantity related to macroscopic,
chaotic motions of the gas, and their related density fluctuations.\\
At small scales, the density field of M1-67 appears remarkably structured in essentially chaotically 
oriented filaments everywhere in the nebula. However, many filaments (especially at the smallest scales) seem to be slightly radially oriented.
This is not surprising, since the most likely force which drives the nebula and the ensuing
turbulence is directed outwards in all directions from the central star.
The presence of filaments
agrees with earlier numerical simulations of turbulent flows ({\it e.g.}
Sreenivasan \& Antonia 1997; V\'azquez-Semadeni 1999, and references therein) or
observations ({\it e.g.} Miville-Desch\^enes {\it et al.} 1999, and references
therein) which have revealed how turbulence is characterized by filamentary
vortical structures and surrounding dissipative sheets. It is worth noting that the empirical PDFs in Figure \ref{levy} are statistically good estimates of the true PDFs in the central parts of the distributions. Clearly, these PDFs are not gaussian at the smallest scales; rather, they are best described by L\'evy stable distributions which are characterized by extended tails. In addition, as the spatial scale increases, the L\'evy index $\alpha$ is found to increase, the gaussian case ($\alpha=2$) being attained for scales above about 4''. This behavior also suggests the presence of vorticity in the flow and is likely related to the phenomenon of intermittency (Lis \etal 1998, and references therein).\\
The velocity field of M1-67 (after correction for bad velocity points related to the largest velocity increments) shows a clear inertial range reminiscent of
turbulence.
Because the ISM and the ejected nebula are
compressible and inhomogeneous, the gas flows being in addition highly supersonic, 
we expect shock wave propagations within M1-67. External forces due to stellar winds
(or gravitation in another astrophysical context like Giant Molecular Clouds: Gill \&
Henriksen 1990) 
play an important role in feeding the ISM with highly pressurized motions,
moving the gas at velocities larger than the sound speed, and ultimately
leading to supersonic turbulence and its related small
scale compressions. Interaction between shock waves
and turbulent fluctuations are expected to distribute the energy {\it dissipation} on a cascade characterized by a steeper law $\zeta(1)=0.5$ (Fleck 1996) than predicted by Kolmogorov's law ($\zeta(1)=1/3$). Our value of $\zeta(1)=0.48$--0.49 strongly suggests that the turbulence traced by the velocity field is compressible. In addition, since our
observations come from a projection on the plane of the sky, we therefore
expect a smearing of the information along each line of sight. This may lead
to additional perturbations in the flows, with apparently fluctuating power-laws as a function of projected spatial separation (O'Dell \& Casta\~neda 1987, and references therein) precluding any inertial regime detection. Figure \ref{vstructure1m167} tells us that this effect is practically negligible, and additionally suggests that the H$\alpha$ layer is thin compared to the projected spatial separations. On the other hand, the value $\zeta(2)=0.90$ is in very good agreement with the empirical Larson-type law which predicts $\zeta(2)=0.8\pm 0.1$ (Larson 1981; Miesch \& Bally 1994) and is associated to fields which are essentially shock-dominated. Note that $\zeta(2)< 2\times \zeta(1)$ results from the slight intermittency of the velocity field, which makes it multifractal rather than monofractal. With $C_1$ close to 0, the process is found to be nearly homogeneous (the fractal dimension of the set contributing to the mean velocity field is $2-C_1\approx 1.96$). On the other hand, the degree of multifractality $\alpha\approx 1.90$--1.92 is close to 2. The latter result suggests that we are far from the so-called monofractal $\beta$-model ($\alpha=0$); rather, the turbulence is best described by the log-normal model ($\alpha=2$) for which high values of the field dominate more than for smaller values of $\alpha$. The estimation of $\alpha$ and $C_1$ for other HII regions appears now essential in order to test whether the multifractal parameters found in the case of M1-67 take on universal values. For comparison with other turbulence studies, one may derive from $C_1$ the standard intermittency parameter $\mu$ which is the autocorrelation exponent of the dissipation field: $\mu=K(2,1)$. For the $\beta$-model ($\alpha=0$), $\mu=C_1$, whereas for the lognormal model ($\alpha=2$), $\mu=2\times C_1$. In our case, with $\alpha\approx 1.91$, we get $\mu\approx 1.90 \times C_1\approx 0.076$.\\
Recall that turbulent velocity measurements in wind tunnel experiments lead to a Kolmogorov-type ({\it i.e.} incompressible; $\zeta(2)=2/3$) turbulence with $\alpha\approx 1.3\pm0.1$, $C_1\approx 0.25 \pm 0.05$ and $\mu\approx 0.35\pm0.1$ (Schmitt \etal 1992). The higher degree of multifractality and the smaller intermittency found for the velocity field of M1-67 is likely a consequence of the compressible nature of the tracing gas. This will be discussed in a forthcoming paper.\\
It is worthy of note that large variations may exist
in the history of WR124's mass-loss and the distribution of the ambient
ISM. This precludes 1) any unique interpretation of the absence of a well-defined
inertial range in the density field; and 2) any unambiguous derivation of the stellar mass-loss history because density fluctuations in the absence of self-gravitational effects are believed to be transient only in the context of supersonic turbulence (V\'azquez-Semadeni 1994). However, note that the marginally inertial range found from the density field is likely a consequence of the rapid regeneration of density structures triggered by the turbulent nature of the velocity field.
Finally, given the extreme perturbations of the velocity and density fields in M1-67, it is virtually impossible to measure
any systematic impact of
the WR (or LBV) wind on the nebular structure.
For whatever reason, any
systematic effects seem to have been randomized such that they have become fully masked. On the other hand, the origin of the $\langle s\rangle$-scale features found from the wavelet analysis of the HST H$\alpha$ image still has to be investigated in detail. These features could be related to either clumps of stellar origin, or smoothed out versions of the bright `bullets' reported in Grosdidier \etal (1998).
On the other hand, the irregular nature of the velocity field is 
likely due to either large variations in the density distribution of the ambient ISM,
or large variations in the central star mass-loss history.\\
On the whole, M1-67 exhibits the signatures of compressible, intermittent turbulence, the driver being ultimately the stellar wind originating in WR124, as opposed to gravity/magnetic fields in the ISM. In addition, from the structure function of order 1 of the density field (which scales as
$\langle|\Delta f(r)|\rangle\sim r^{0.76}$ at the smallest scales), we infer the fractal dimension of M1-67 to be
$D\approx$ 2.2--2.3. This result is very close to the values often quoted for molecular clouds and indeed is essentially the result found in the wavelet
analysis of Gill \& Henriksen (1990). Therefore, although different drivers may be related to the generation of supersonic, compressible turbulence, it appears that the resulting phenomenology essentially remains the same.

\acknowledgments
The authors are grateful to the anonymous referee for stimulating
and constructive criticism.
YG acknowledges financial aid from the French Ministry of
Foreign Affairs. AFJM and GJ are grateful to NSERC
(Canada) and FCAR (Qu\'ebec) for financial support. AFJM acknowledges
the award of a Killam Fellowship from the Canada Council for the Arts.

\clearpage

\clearpage

\end{document}